\documentclass{PoS}

\newcommand{\beq}{\begin{equation}}
\newcommand{\eeq}{\end{equation}}

\newcommand{\Xmumax}{X^{\mu}_{\rm max}}
\newcommand{\Xmax}{X_{\rm max}}

\newcommand{\nmu}{N_\mu}

\newcommand{\lnnmu}{\ln\!\nmu}
\newcommand{\mlnnmu}{\langle \ln\!\nmu \rangle}

\newcommand{\xmax}{X_{\rm max}}

\title{Probing High-Energy Hadronic Interactions with Extensive Air Showers}

\ShortTitle{Probing High-Energy Hadronic Interaction with EAS}

\author{\speaker{Lorenzo Cazon}\\
        LIP, Av. Prof. Gama Pinto 2, 1649-003 Lisbon, Portugal\\
        E-mail: \email{cazon@lip.pt}}

\abstract{In this article, I will review how observations performed with extensive air showers are being used to probe hadronic interactions at high energy. I will briefly overview the new studies exploring the connection between the dynamics of air showers and multiparticle production,  and how this knowledge can be translated into constraints on high energy hadronic models. I will also overview direct measurements, complementary to, and beyond the reach of, accelerator experiments. 
}

\FullConference{36th International Cosmic Ray Conference -ICRC2019-\\
		July 24th - August 1st, 2019\\
		Madison, WI, U.S.A.}

\begin{document}
\setcounter{page}{2}

\section{Introduction}

Above $10^{15}$ eV, cosmic rays (CR) are detected by means of Extensive Air Showers (EAS).
EAS encode information about the primary CR, (energy, mass number, and arrival direction), and also details of the particle interactions therein, many of them occurring at phase-space regions well beyond the reach of accelerators. 
A centre-of-mass energy of $\sqrt{s}\simeq 14$ TeV corresponds to an energy of $10^{17}$ eV in the laboratory frame, where we usually speak about Ultra High Energy Cosmic Rays (UHECR). UHECR reach up to $10^{20}$eV ($\sqrt{s}\simeq 450$ TeV), which is $\sim 30$ times the energy achieved in the LHC. 

High energy interaction models extrapolate our knowledge from accelerator measurements  to the forward and ultra high energy region, with an uncertainty that increases as we depart from the tested regions. The interpretation of EAS observables in terms of the UHECR mass inherits the model uncertainties as one of the main contributions to the systematic uncertainties. In this sense, primary mass and hadronic interactions uncertainties are difficult to untangle from an experimental point of view.

Is it possible to break this degeneracy? Testing particle physics mostly relies on comparisons of full EAS simulations with data taken by EAS experiments. These simulations include all interactions occurring within the air-shower at high and low energies. The secondaries arising after the first interaction undergo successive particle reactions creating again new particles, and making EAS achieve macroscopic size. As a consequence, most details of the multi-particle production of the first interaction are hidden within the vastness of all particles of the cascade. The most common approach is to look at EAS and to make a direct comparison  with  simulations. Each possible primary mass admixture of UHECR corresponds to a region of the $n$-dimensional phase-space of observables when simulated with a given model. If one simply allows all possible mass combinations, one would obtain the total allowed phase-space for each model.  Data must fall within the allowed region for a model to be valid, or in other words, all observables must be consistent for a given mass composition. If data fall off the allowed observable phase-space, inspiration might occur by  \emph{reverse-engineering} solutions by playing with some of the knobs provided by the high-energy interaction models, (without violating available accelerator data), or sometimes resorting to include new phenomena, until EAS observations are well accommodated within models.

EAS are measured in two main ways: by collecting the electromagnetic radiation emitted after the cascading charged particles (mostly electrons) interact with air molecules and the geomagnetic field, (UV-light: Fluorescence light, Cherenkov, MHz radio: Geosyncrotron, Cherenkov), and  by detecting the particles that reach the ground level (mostly electrons, photons and muons).

In Section 2, we will revisit our understanding of the air shower, and its relation with the microscopic variables provided by the hadronic models. In Section 3, we will quickly review the EAS observables that constrain hadronic models and we will discuss their intepretation in Section 4. In Section 5, we will just refer to those EAS observables directly linked to microscopic variables. The conclusions will be given in Section 6.

\section{Air Shower Physics}

  
\begin{figure}[h]
  \begin{minipage}{\textwidth}
  \centering
  \includegraphics[width=0.30\textwidth,trim=20 10 00 5,clip]{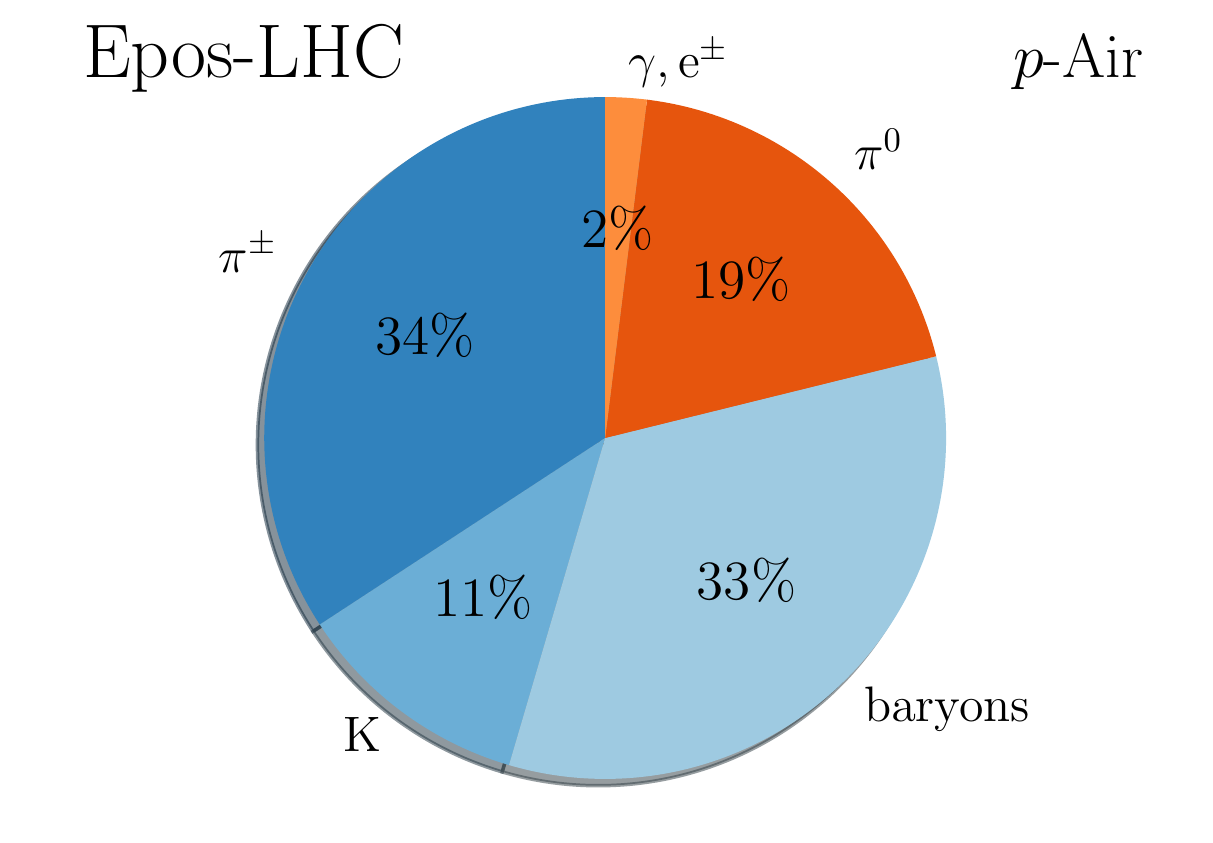}
 \includegraphics[width=0.30\textwidth,trim=20 10 00 5,clip]{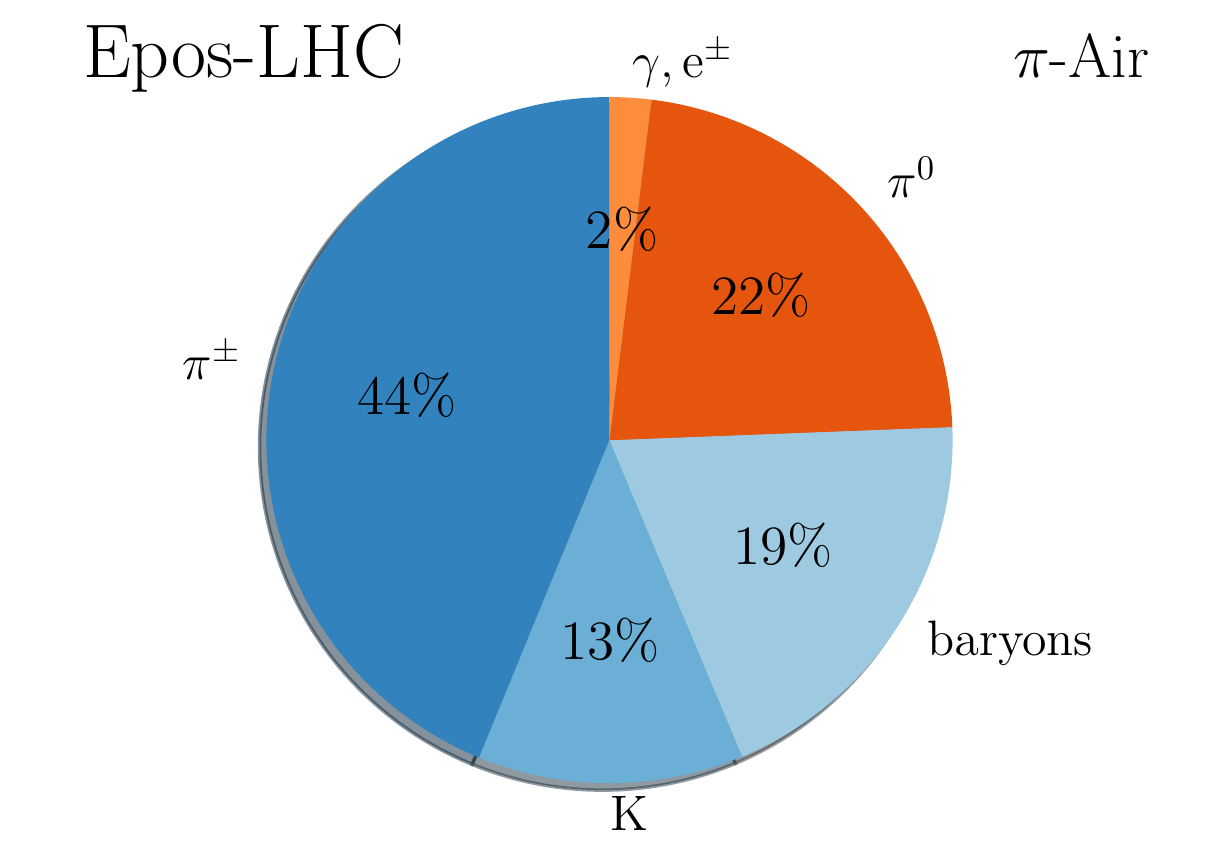}
 \includegraphics[width=0.30\textwidth,trim=20 10 00 5,clip]{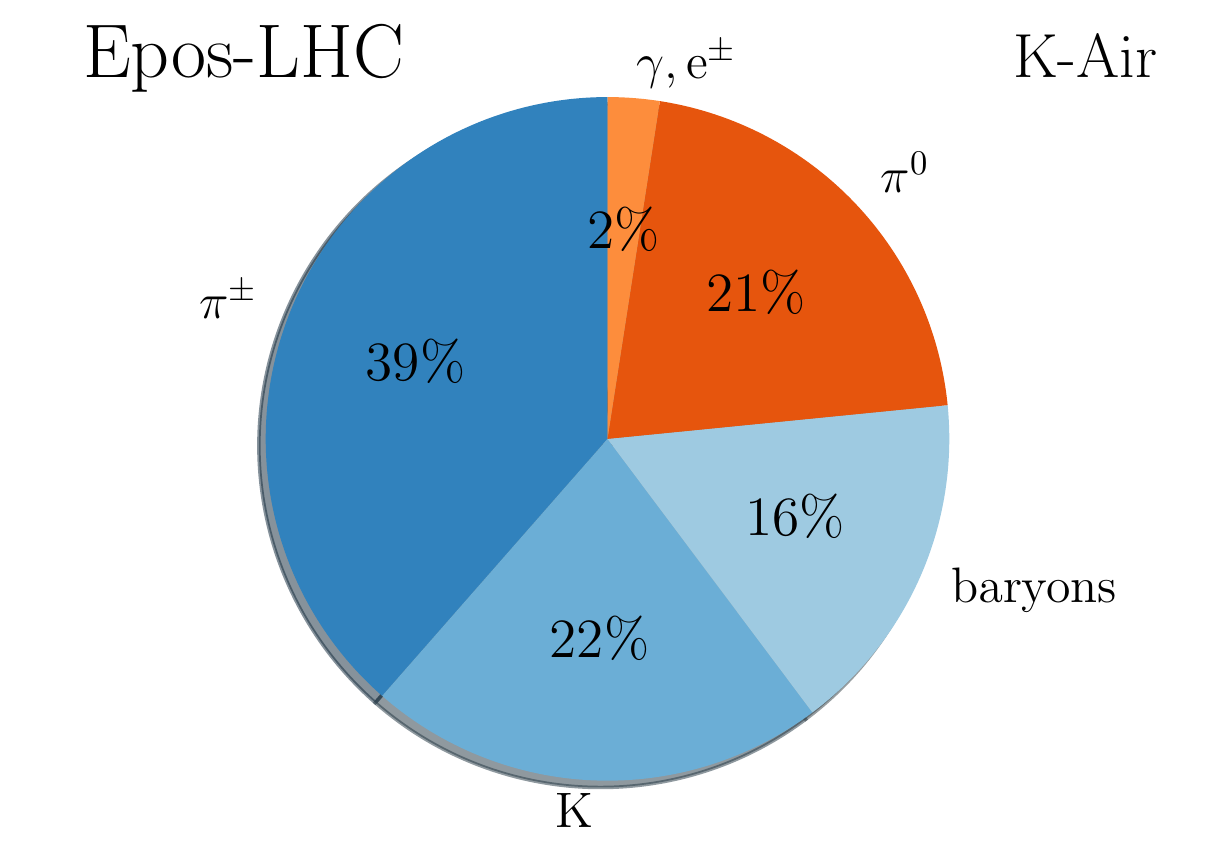}
\end{minipage}
  \begin{minipage}{\textwidth}
     \centering
    \includegraphics[width=0.15\textwidth,trim=25 0 30 0,clip]{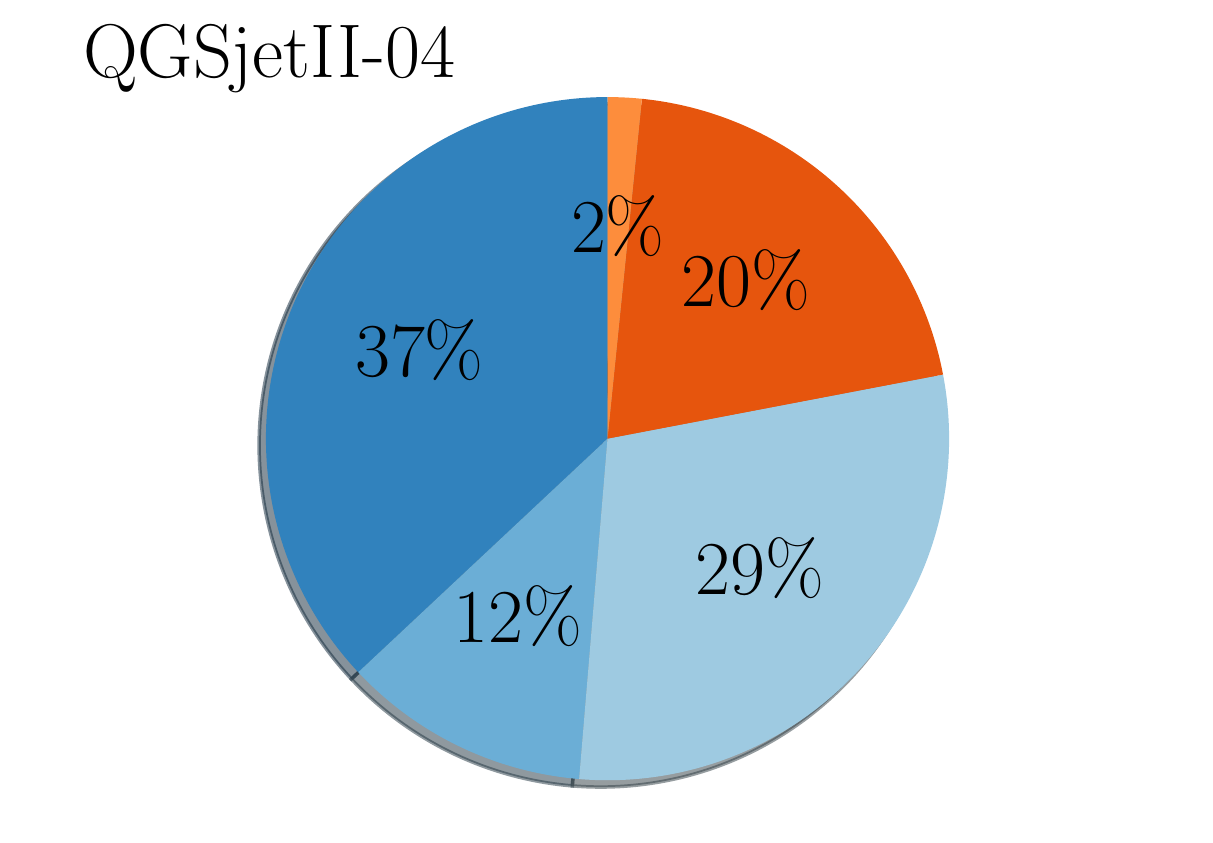}
    \includegraphics[width=0.15\textwidth,trim=25 0 30 0,clip]{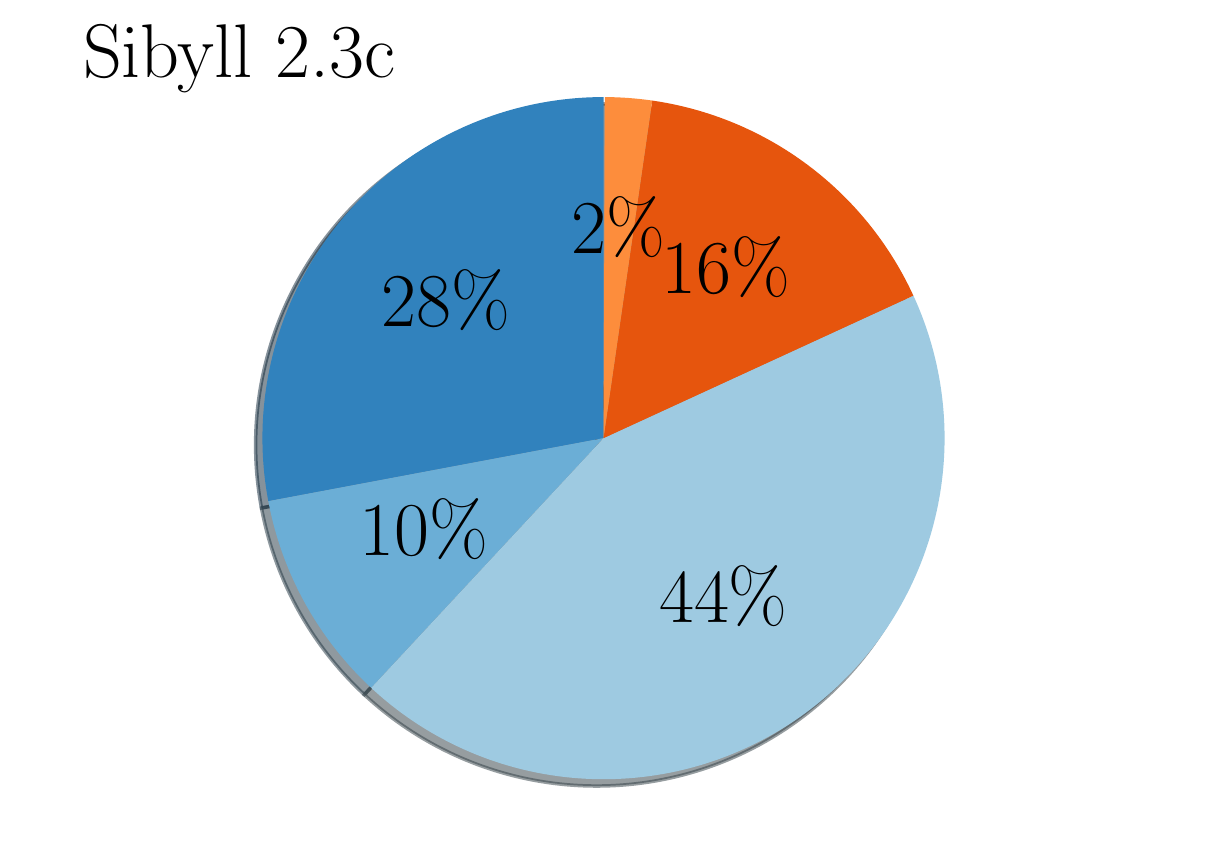}
      \includegraphics[width=0.15\textwidth,trim=25 0 30 0,clip]{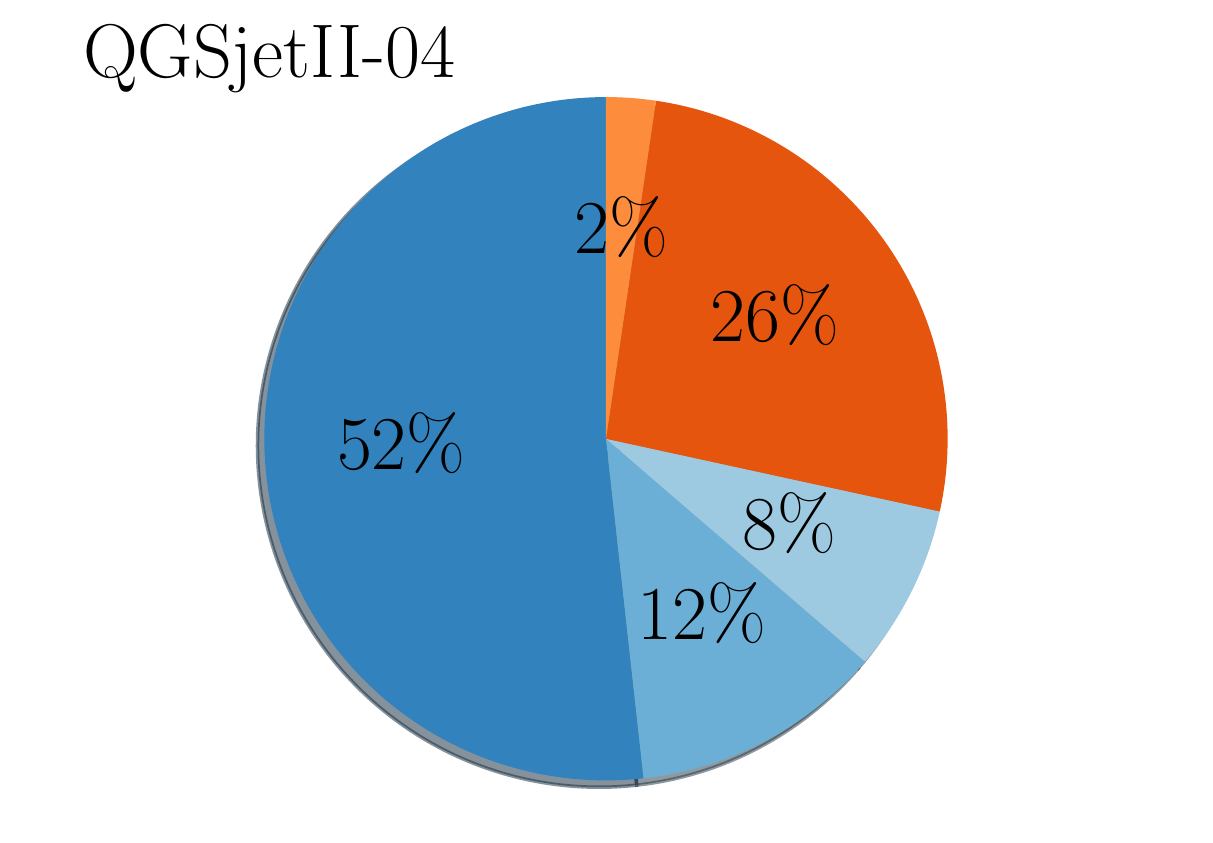}
    \includegraphics[width=0.15\textwidth,trim=25 0 30 0,clip]{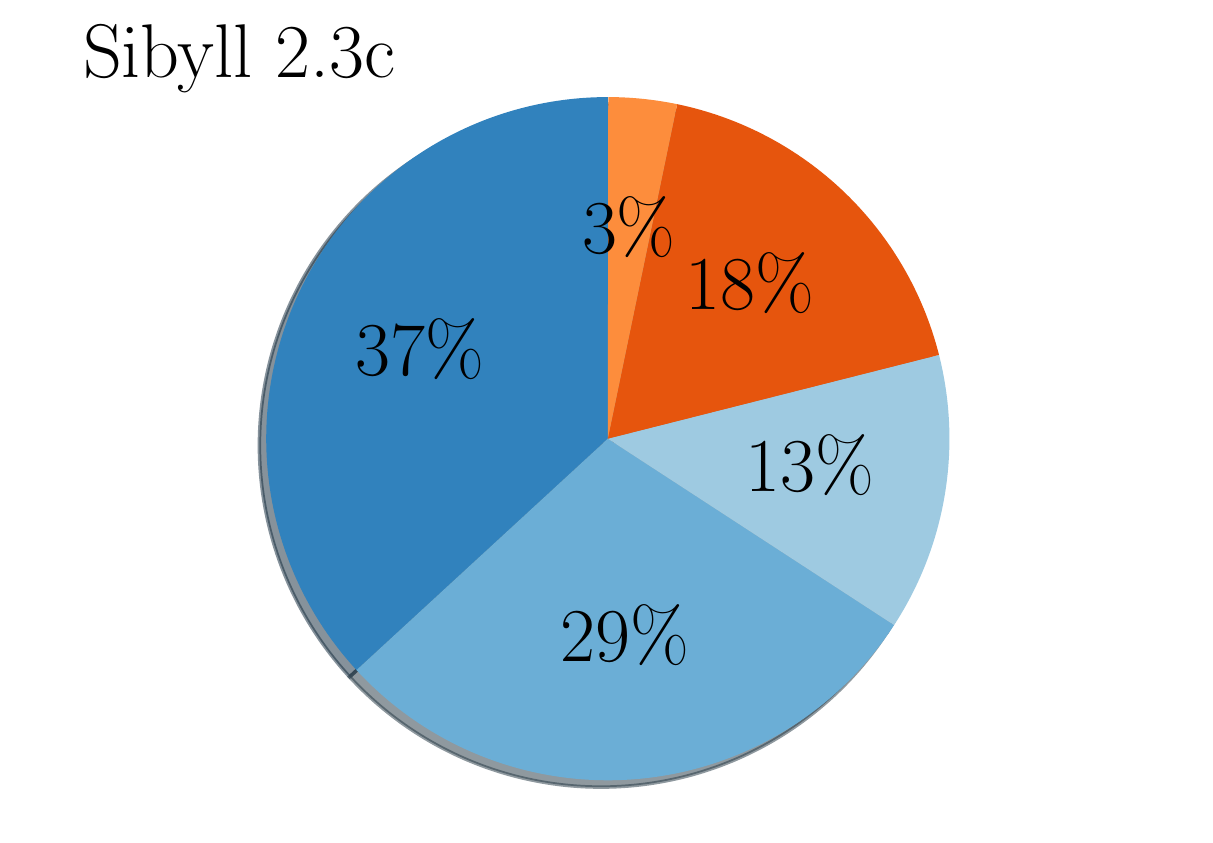}
      \includegraphics[width=0.15\textwidth,trim=25 0 30 0,clip]{fig-en-share-qgsjetII04-fluka-piplus-no-label.pdf}
    \includegraphics[width=0.15\textwidth,trim=25 0 30 0,clip]{fig-en-share-sibyll235-fluka-kplus-no-label.pdf}    
   \end{minipage} 
  \caption{Average share of energy among different groups of
particles in $p$-Air (left), $\pi$-Air (centre), and $K$-Air (right) interaction at $10^{19}$ eV, simulated with different models, as labelled. Numbers are in percent. Particles contributing to the electromagnetic component are shown in shades of red, particles in the hadronic component are shown in shades of blue. The contributions to the $\pi^\pm$ sector include the decay $\rho^0 \rightarrow \pi^+ \pi^-$. The baryon sector includes $p, n, \Lambda$ and their antiparticles.
  }
  \label{fig-en-share}
\end{figure}

\begin{figure}[h]
  \begin{center}
    \includegraphics[width=0.350\textwidth,clip, trim=20 5 40 35]{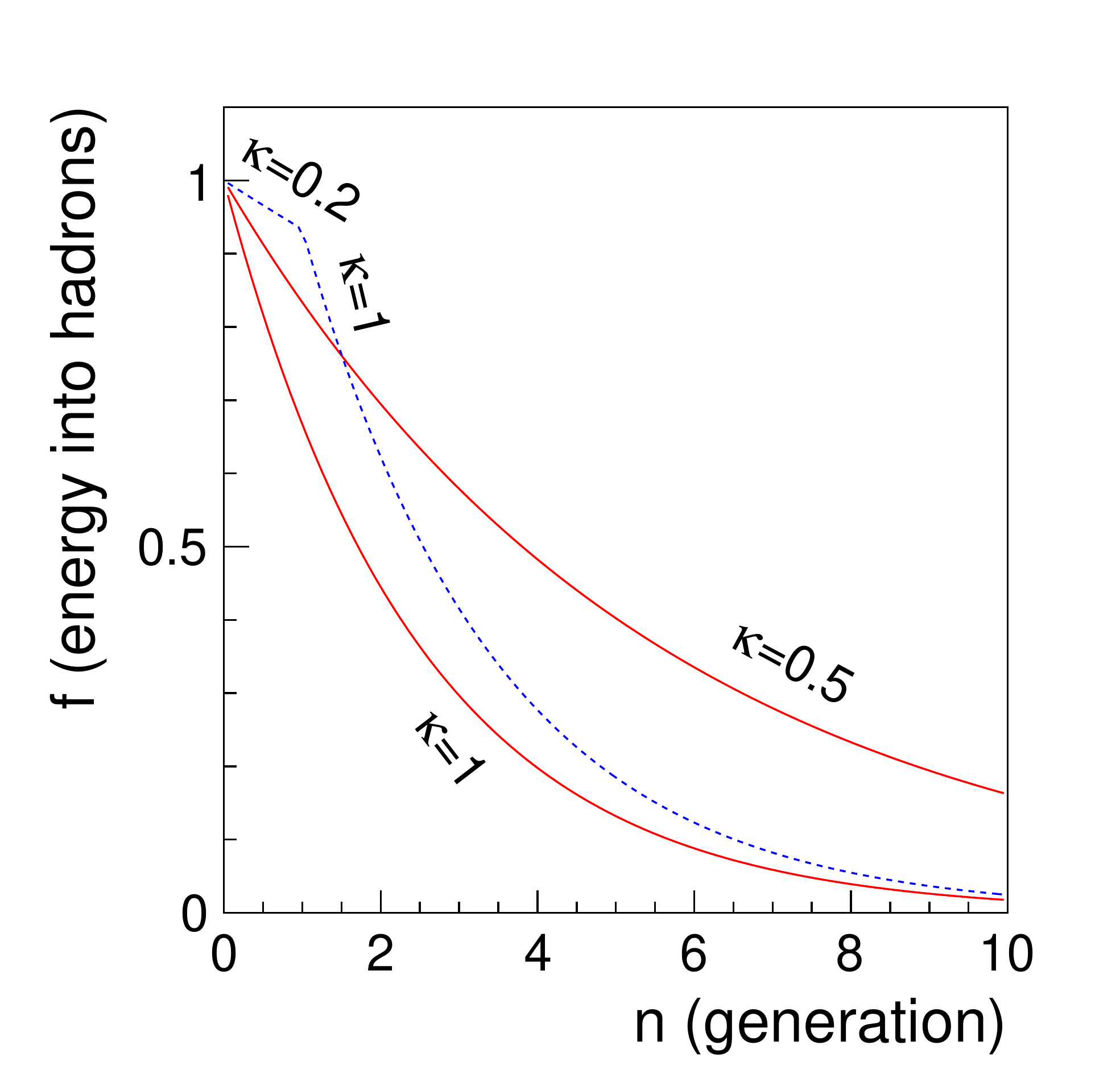}
    \includegraphics[width=0.550\textwidth,trim=00 05 00 00,clip]{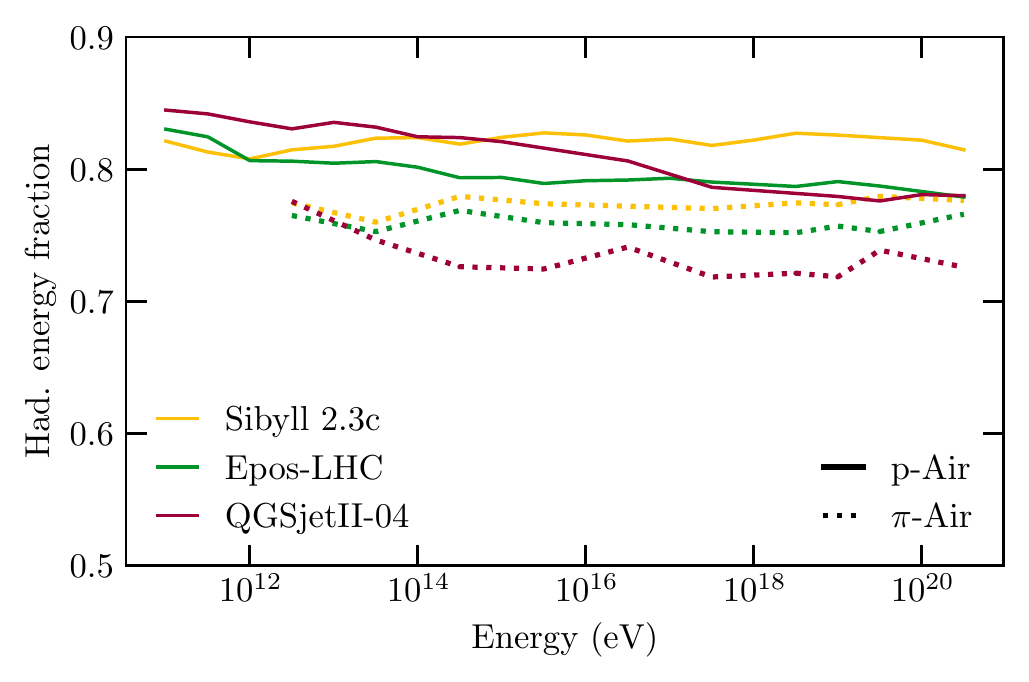}
    \caption{
  Left: energy fraction evolution with generation $n$, taken from a Heiler-Mattews model with different $\kappa$ parameters, taken from \cite{Cazon:2013dm}. Right: Hadronic energy fraction $f$ as a function of energy, for $p$-Air and $\pi$-Air interactions simulated with several models. 
}
\label{fig:f}       
\end{center}
\end{figure}

After the first UHECR-air interaction, approximately $\sim 75\%$ of the energy 
goes into secondary  mesons 
and baryons, which continue interacting,
creating the  so-called {\it hadronic cascade}. Figure \ref{fig-en-share} shows
the average share of energy among different groups of
particles in $p$-Air (left), $\pi$-Air (centre), and $K$-Air (right) interaction at $10^{19}$ eV, simulated with different models. The contributions to the $\pi^\pm$ sector include the decay $\rho^0 \rightarrow \pi^+ \pi^-$, and the baryon sector includes $p, n, \Lambda$ and their antiparticles. When the average energy per meson decreases, it eventually becomes more likely that mesons decay rather than interact. This is called critical energy, and it is found  to be: $\xi^{\pi^\mp}_{\rm crit}={\cal O}$(100~GeV), $\xi^{K^\pm}_{\rm crit}={\cal O}$(1000~GeV), $\xi^{K^0_L}_{\rm crit}={\cal O}$(200~GeV), $\xi^{K^0_S}_{\rm crit}={\cal O}$(30~TeV).
This is the stage where most muons are formed.  Muons are the main messengers from the hadronic cascade.

The {\it Electromagnetic (EM) cascade} consists of photons and electrons. It is fed from the hadronic cascade by the decay of neutral pions $\pi^0$ into photons, which then keep multiplying in number by pair production and bremsstrahlung.  In each interaction, approximately $\sim 25\%$ of the energy is transferred from the hadronic to the EM cascade by $\pi^0$ decay, being the largest absolute contribution the one from the first interaction. In Figure \ref{fig-en-share}, the energy from particles contributing to the electromagnetic component are shown in shades of red, particles in the hadronic component are shown in shades of blue.

The most relevant features of a hadronic shower can be approximately described by a pionic Heitler-Matthews model \cite{Matthews:2005sd}: in each hadronic generation\footnote{Generation $n=1$ corresponds to the particles emerging from the first interaction, generation $n=2$, corresponds to the particles emerging from the interaction of particles of generation $n=1$, and so on so forth.} $n$, charged and neutral pions are created in a $f=2/3$ and $f_{\rm EM}=1/3$ proportion.  The energy fraction carried by the sum of all charged pions in generation $n$ to the total shower energy $E_0$ is
\begin{equation}
\frac{\sum E_{\pi}}{E_0}=f^n=\left(1-f_{\rm EM}\right)^{n}
\end{equation}
 In each generation, the energy carried by charged pions $\sum E_{\pi}$ is reduced by a factor $f$.  In a more realistic approach, we can include an effective factor $\kappa$ that modifies the amount of energy going to $\pi^0$.  For instance, if a leading baryon takes $(1-\kappa)E_0$, $\kappa$ accounts for the inelasticity, and the energy flowing to the EM cascade is reduced as $f_{\rm EM}=1/3 \kappa$
  as explained in \cite{Matthews:2005sd}. There might be other mechanisms that could effectively reduce the feeding to the EM channel, for instance, increasing the amount of kaon production \cite{AlvarezMuniz:2012dd}.

The energy share between both cascades evolves with the hadronic generation as shown in Figure \ref{fig:f} (left) for the pure pionic cascade ($\kappa=1$) and a more realistic case ($\kappa=0.5$, $f_{\rm EM}=0.17$)\cite{Cazon:2013dm}. In the beginning all the energy is in the hadronic sector. After 3  generations, most of the energy has gone to the EM cascade, and the hadronic and EM cascade can be considered decoupled. The number of muons arising from hadronic cascade is approximately $\nmu=E_0 f^{c} / \xi^{\pi^\mp}_{\rm crit}$, where all successive interactions down to the critical generation $n=c$ have contributed. On the other hand, the EM cascade is dominated by the most energetic contributions.
 
\subsection{Electromagnetic and hadronic Shower}

Beyond the so called   {\it pure muon component} (directly emerging from the hadronic cascade),  and the {\it pure EM component}, (from high energy $\pi^0$ decays), one can distinguish other contributions \cite{Ave:2017uiv} \cite{Ave:2017wjm}:
  {\it  EM from muon decay} or {\it muon halo} which stems from the decay of muons, and therefore scales with the hadronic component of the shower;
  {\it  EM from low energy $\pi^0$ decay} which is a small contribution to the EM cascade but nevertheless is coupled with the hadronic cascade;
  and {\it  muon from photo-production}, which stems from the pion production after photon-air interactions, and is therefore coupled to the EM cascade.

Photon and electron initiated showers are well studied in literature. The longitudinal development of the number of electrons is described by a Greisen profile. For a hadron induced shower, each $\pi^0$ decay initiates its own contribution to the electromagnetic component, resulting into a Gaisser-Hillas profile in the longitudinal component.
In \cite{Smialkowski:2018reh,Giller:2014uja,Nerling:2005fj,Giller:2004cf} an analytical description for the electron energy spectra, the lateral distributions, and the angular distribution were shown to be universal if properly expressed in terms of shower age and Moliere radius. The idea of universality of showers is crucial from an experimental point of view, as it allows one to fit universal templates to observed particle distributions to reconstruct the fundamental degrees of freedom of the shower.

\begin{figure}[!ht]
  \begin{center}
    \includegraphics[width=0.31\textwidth,trim=07 07 05 07, clip ]{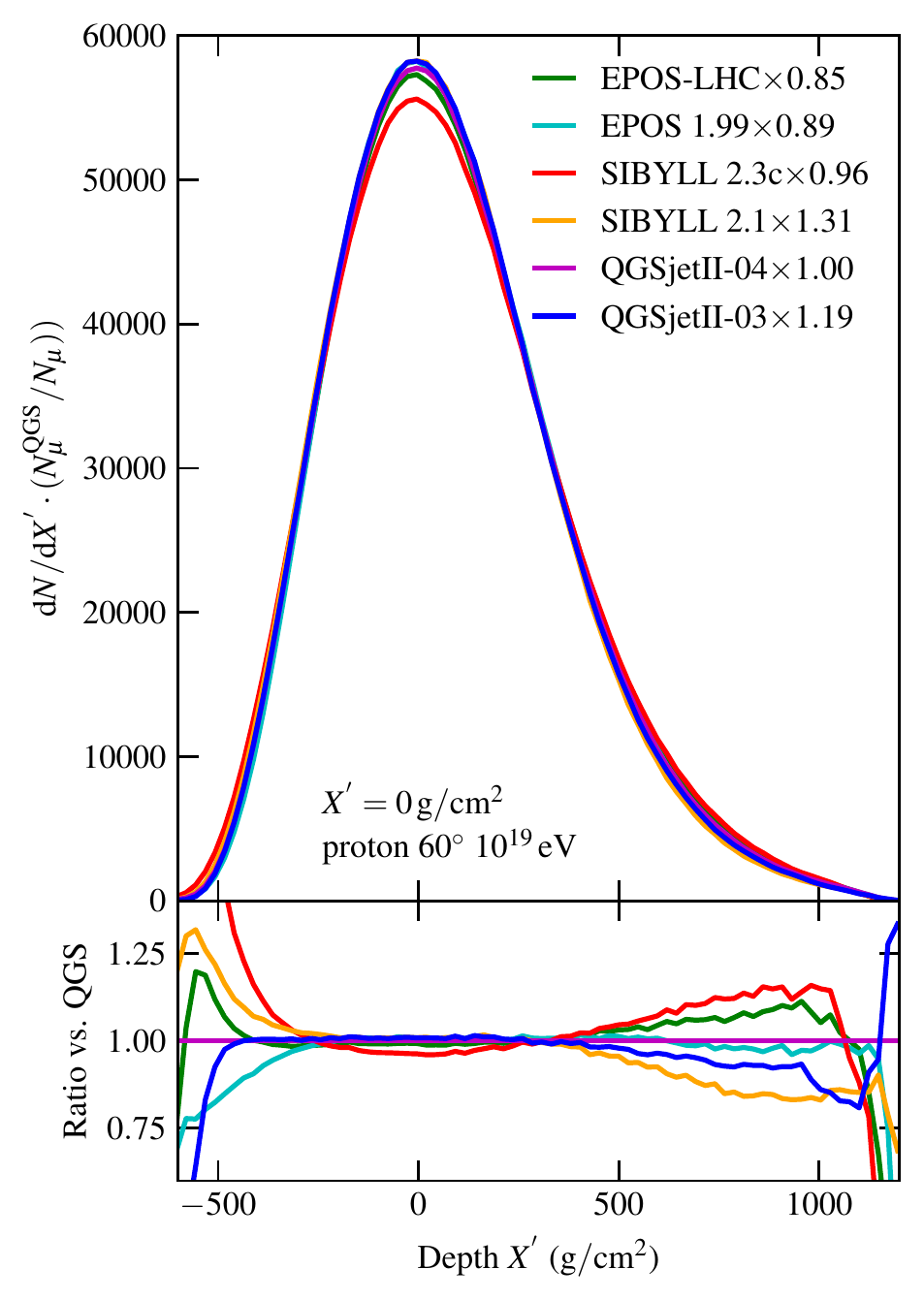}
    \hfill
    \includegraphics[width=0.31\textwidth,trim=07 07 05 07]{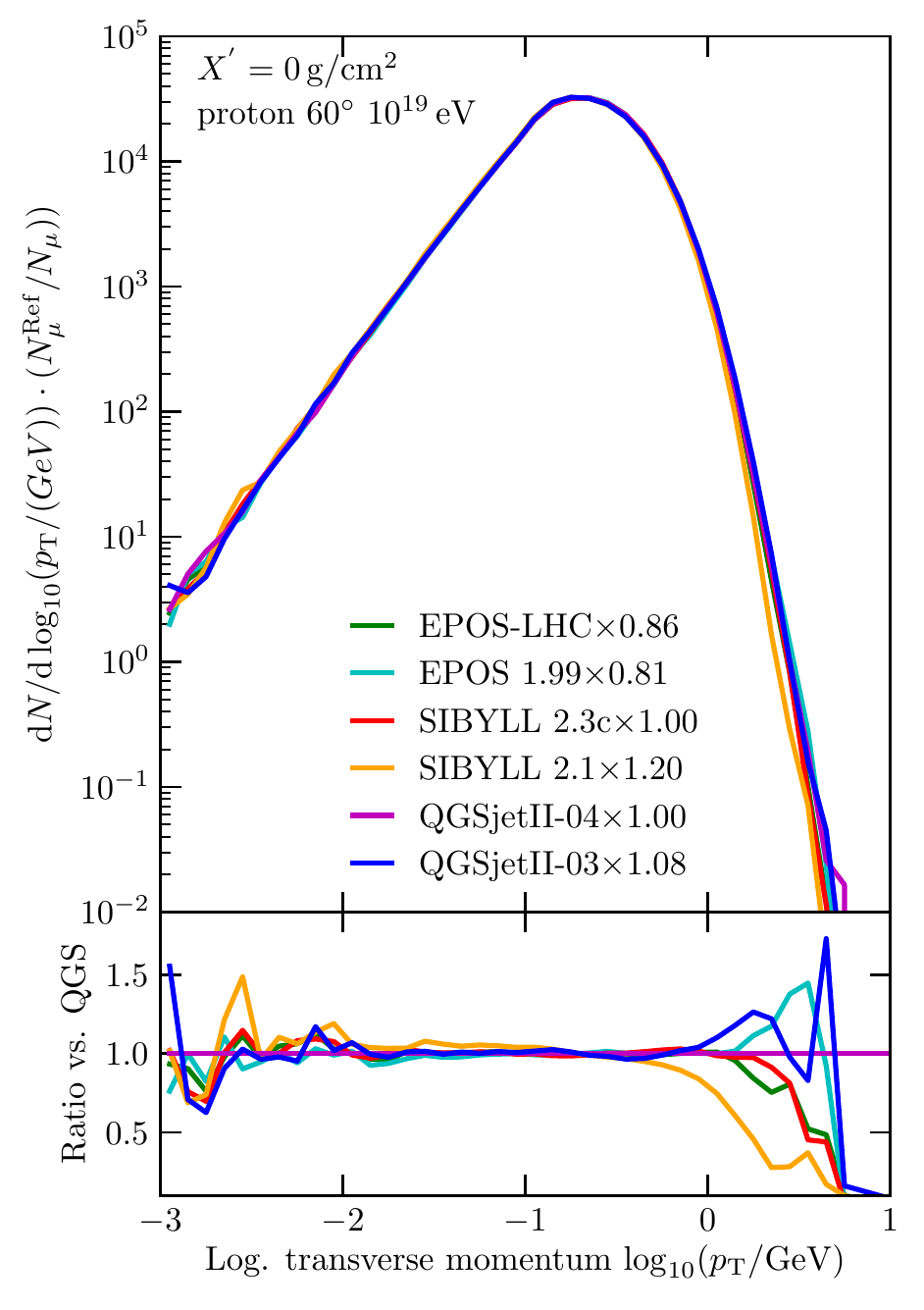}
    \hfill
    \includegraphics[width=0.31\textwidth,trim=07 07 05 07]{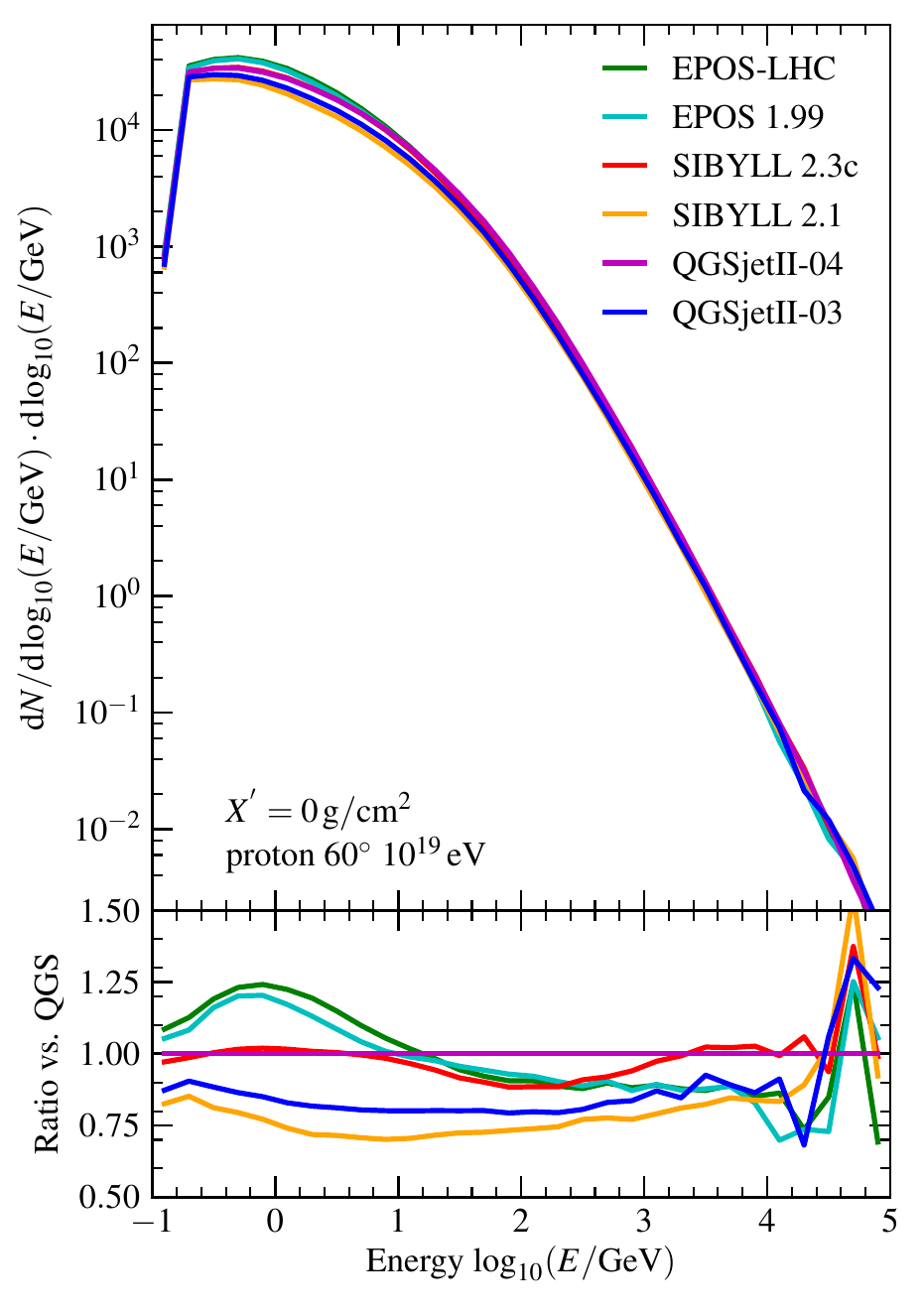}
    \caption{{\it Total/true} Muon Production Depth distribution (left),
      transverse momentum distribution at production (centre),
      muon energy distribution at production (right),
     for proton showers simulated at $10^{19}$ eV, taken from \cite{Cazon:2019nop}.}
    \label{f:universality}
  \end{center}
\end{figure}

In \cite{Cazon:2012ti} it was shown that the muon component can be fully described by the function
\beq
\frac{d^{3} N}{dX\, dE_i\,dcp_{t}}={\cal \nmu} \, f(X-{\cal \Xmumax},E_i,cp_{t})
\label{eq:mu}
\eeq 
where  ${\cal \nmu}$ is the total number of muons produced in the shower, $X$ is the slant depth, ${\cal \Xmumax}$ is the depth where the rate of muon production reaches a maximum, $E_i$ is the energy of the muons at production and   $cp_t$ is the transverse momentum with respect to the shower axis. In  \cite{Cazon:2012ti} it was demonstrated that this distribution can be used to propagate muons  to obtain any distribution at ground: lateral distribution function, {\it apparent} Muon Production Depth (MPD) distribution and its  maximum $\Xmumax$, arrival time distribution, energy spectrum, {\it et cetera}. The 3-dimensional distribution of Eq. \ref{eq:mu} is directly inherited from the hadronic cascade. The  $cp_t$-distribution and to a lesser extent the {\it total/true} MPD distribution  are universal across primaries and models when referred to the maximum $X' \equiv X-{\cal \Xmumax}$, whereas the $E_i$-distribution shows sizeable differences across models and primaries \cite{Cazon:2012ti,Cazon:2019nop}. Figure  \ref{f:universality} shows 3 of these distributions for the post-LHC hadronic models.


Muons exit the shower axis with an angle $\gamma$ determined by the energy and transverse momentum of the parent pion production ($\sin(\gamma)= \frac{cp_t}{E_i}$). Pions decay after travelling $c \tau=E_i/(m_\pi c^2)  c \tau_0$.  The perpendicular distance to the shower axis before pion decay is $r_{\pi}=c \tau \sin(\gamma)=cp_t c \tau_0/(m_\pi c^2)$. The $p_t$-distribution of muons (and the decaying pions) can be described by \linebreak
$dN/dp_t = (p_t/Q^2)\exp(-p_t/Q)$, with $cQ \sim 0.2$ GeV. It follows that 60\% of muons are produced within $r_\pi<22$ m, much smaller than the typical lateral distances of observation in experiments. Therefore, in most cases one can use the approximation that muons are produced in the shower axis.


\subsection{High Energy Hadronic Models}

\begin{figure}[!ht]
  \begin{center}
    \includegraphics[width=0.450\textwidth]{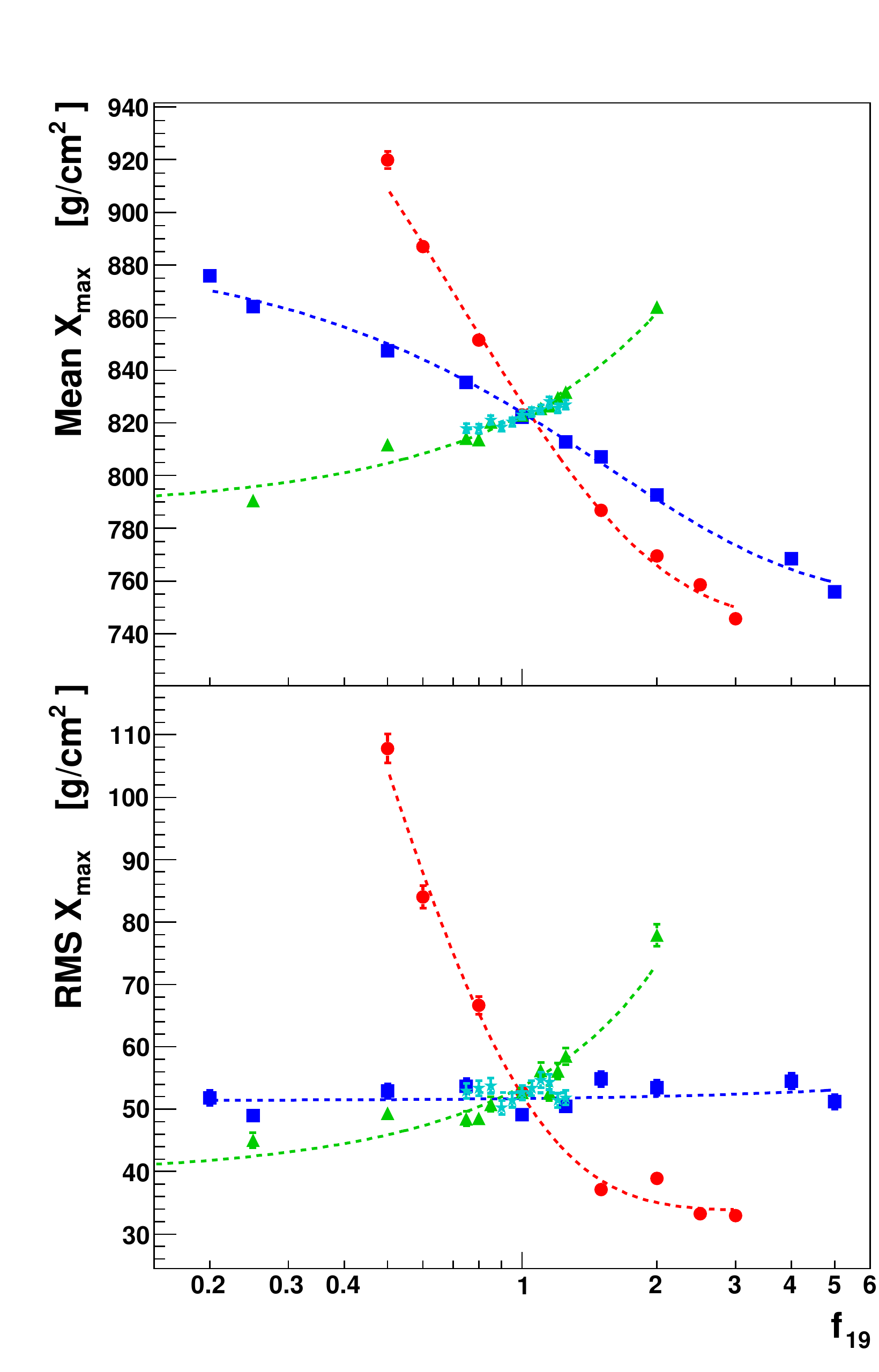}
    \hfill
    \includegraphics[width=0.450\textwidth]{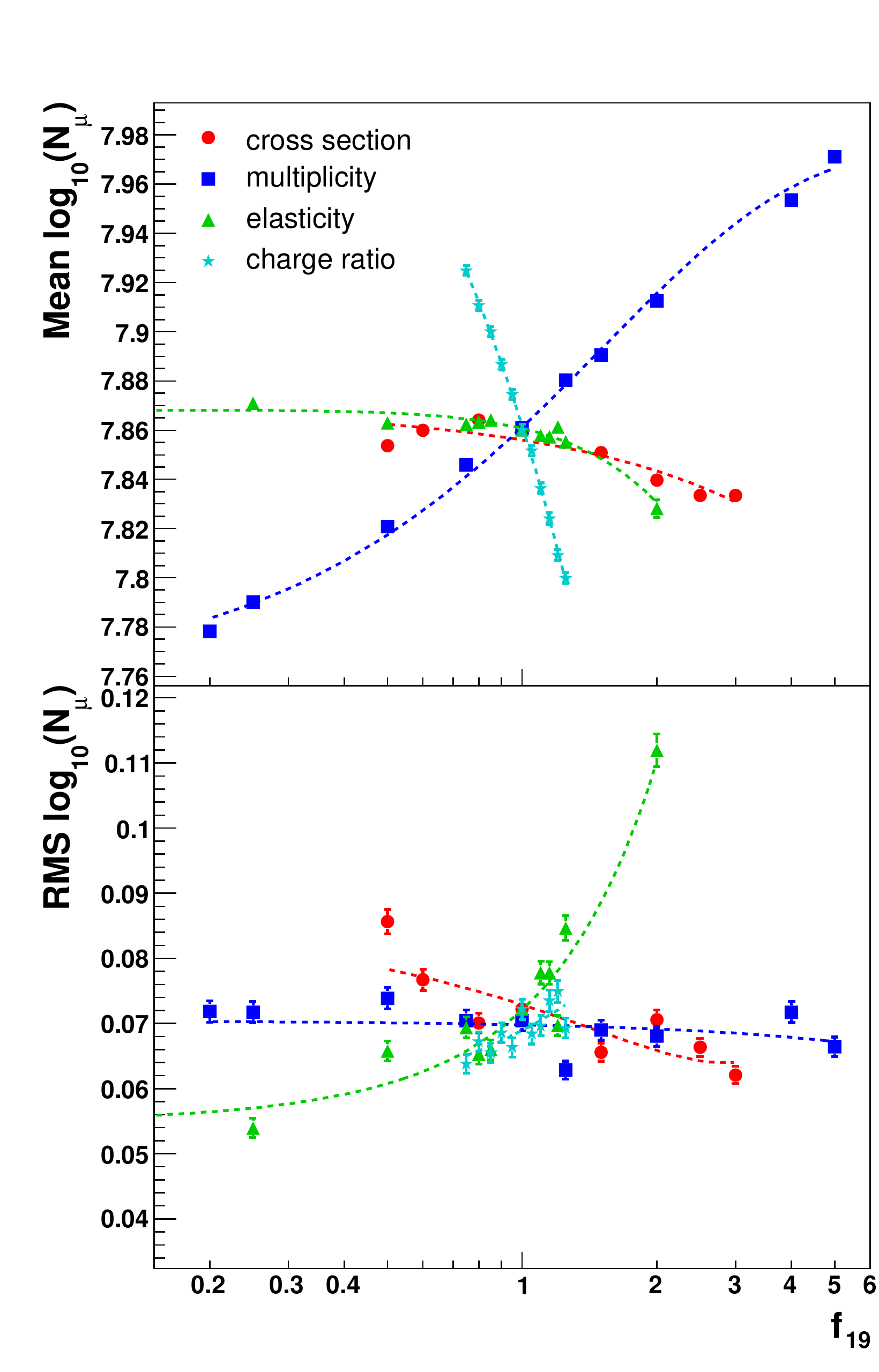}
    \caption{Impact of a modified extrapolation of hadronic interaction features on depth of shower maximum $\Xmax$ (left) and the number of muons $\nmu$ (right) as a function of the scaling hadronic parameter $f_{19}$ for a simulated proton shower at $10^{19}$ eV with SIBYLL-2.1.
(Taken from \cite{Ulrich:2010rg}.)
    }
     \label{f:ulrich}
  \end{center}
\end{figure}

The impact of different extrapolations of some  hadronic interaction parameters on the predictions of several air shower observables was studied in \cite{Ulrich:2010rg}. These parameters were the p-Air cross-section, the total multiplicity, the elasticity (fraction of energy taken by the leading particle) and the charge-ratio, defined as $n_{\pi^0}/(n_{\pi^0}+n_{\pi^+}+n_{\pi^-})$ and which might be taken as a proxy of $f_{\rm EM}$.  The model used a continuous and smooth evolution of hadronic particle production from  $10^{15}$eV to high energy, and used the scale $f_{19}$ as scaling factor at $10^{19}$ eV. In Figure \ref{f:ulrich}, left-bottom panel, one can see the effects on the fluctuations of the longitudinal development of the EM component (RMS$(\Xmax)$), which depends mainly on the cross section and less strongly on the elasticity.  The muon number (right, upper panel) can be increased by increasing the total multiplicity, or by decreasing the charge-ratio.

If one asks now what can be actually measured in accelerator experiments,
the most easily accessible region is around the central pseudorapidities, which also contains the highest rapidity-density of produced particles.
Nevertheless,   the forward region is the one which carries most of the energy after the collisions. This is where the cascading process occurs and new particles are created within the EAS.  LHC first data were compared to hadronic models used in UHECR in \cite{dEnterria:2011twh}, and there is work in a variety of forward detectors to study the energy spectra of forward particles \cite{N.Cartiglia:2015gve}, which have a direct impact on air shower development.
We will focus in three hadronic interaction models, which are commonly used to simulate EAS, and were updated to take into account LHC data at 7 TeV: QGSJETII-03
\cite{
  Ostapchenko:2006ii}
updated into QGSJETII-04 \cite{Ostapchenko:2011,Ostapchenko:2013pia}, EPOS 1.99
\cite{
  Pierog:2009}
updated to EPOS-LHC \cite{Pierog:2015}, and SIBYLL-2.1
\cite{
  Ahn:2009}
updated to SIBYLL-2.3 \cite{Riehn:2015} and SIBYLL-2.3c~\cite{Riehn:2017mfm}.

\begin{figure}[!ht]
  \begin{center}
    \includegraphics[width=0.45\textwidth]{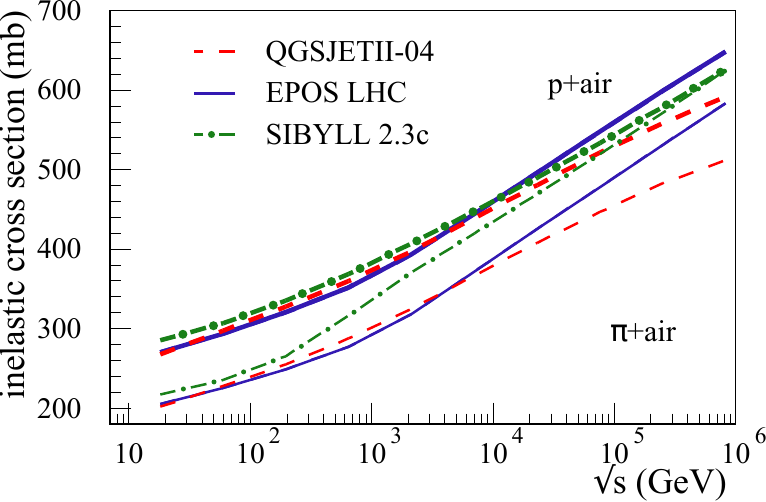}
    \hfill
    \includegraphics[width=0.45\textwidth]{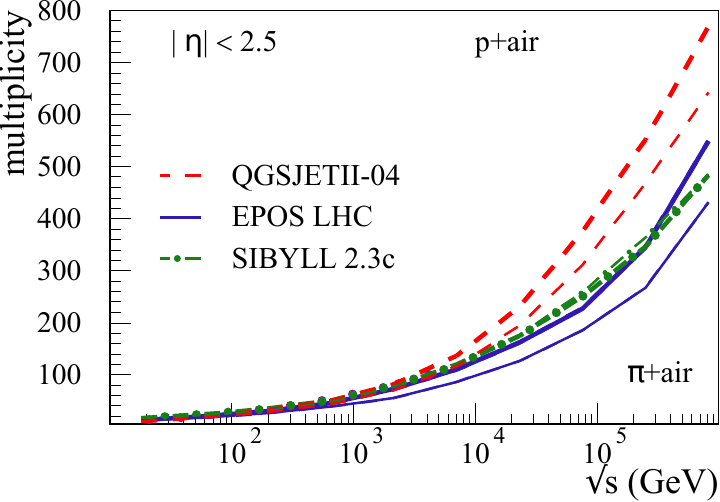}
    \caption{Inelastic cross sections (left) and multiplicity for $|\eta|<2.5$ (right) for p-air (thick lines) and $\pi$-air (thin lines). (Taken from \cite{Pierog:2019srb}.)
    }
    \label{f:pAir}
  \end{center}
\end{figure}

The inelastic cross section determines the depth of the first interaction $X_0$, as well as the rate of interactions of the secondary particles. It has a direct impact in the distribution of $\Xmax$, as can be seen in Figure \ref{f:ulrich}. The p-p cross section is very well described up to the LHC energy, and the extrapolations up to the highest energies is very similar between models \cite{dEnterria:2016oxo}. However, there are some differences in the extrapolations of the p-air and $\pi$-air inelastic cross-sections as shown in Figure \ref{f:pAir}, which are more relevant for the case of $\pi$-air interactions, the most numerous reactions along the shower development. The average multiplicity is plotted in Figure  \ref{f:pAir} (right), where differences appear between models at the highest energies. So, for both cross section and multiplicity, the extrapolation to the highest energy 
in nuclear and pion interactions is still uncertain because of the lack of data at high energy and with light ions.


\begin{figure}[!ht]
  \begin{center}
    \includegraphics[width=0.45\textwidth]{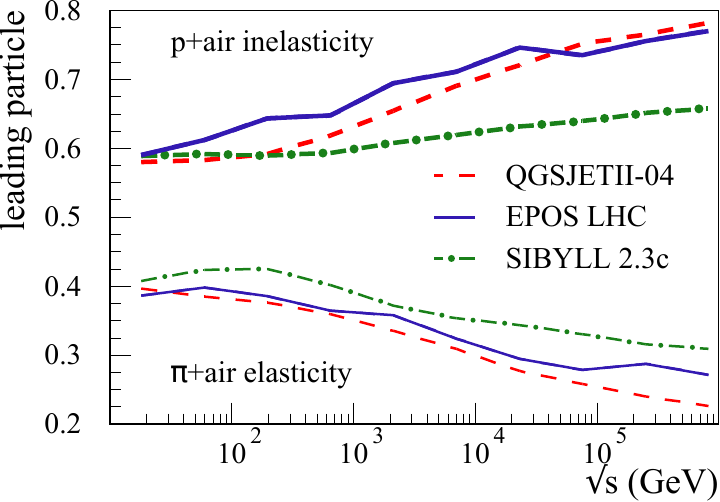}
    \hfill
    \includegraphics[width=0.45\textwidth]{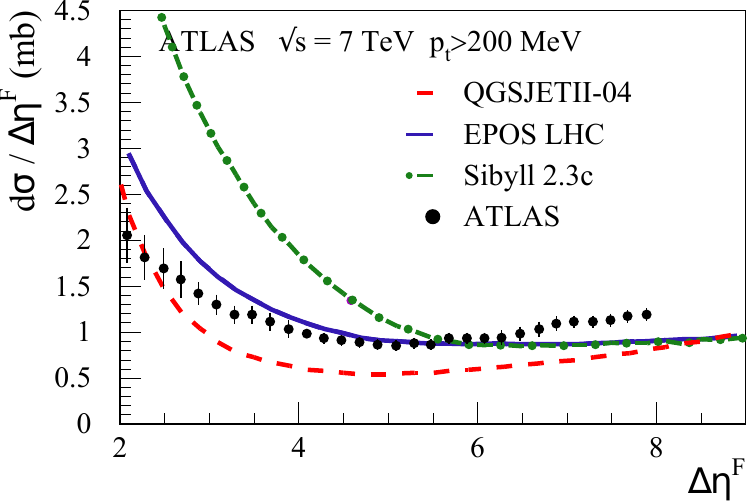}
    \caption{Left: Inelasticity in p-Air interactions (thick lines) and elasticity for $\pi$-Air interactions (thin lines). Right: ATLAS measurement of the pseudorapidity gap $\Delta \eta^F$ for particles with $p_{t,cut} >$ 200 MeV in minimum bias events at 7 TeV.  (Taken from \cite{Pierog:2019srb}.)
    }
    \label{f:elasticity}
  \end{center}
\end{figure}

The elasticity is only indirectly constrained by collider experiments. Figure  \ref{f:elasticity} (left) displays the extrapolations for p-air and $\pi$-air interactions, showing a  large spread in the models.
Figure  \ref{f:elasticity} (right) displays the rapidity gap cross-section (range in pseudorapidity without particle detection), showing discrepancies between models and data. Large rapidity gaps come from single diffraction events. In general, elasticity and rapidity gap distributions show that models display sizeable uncertainties describing diffractive events.


\begin{figure}[!ht]
  \begin{center}
    \includegraphics[width=0.450\textwidth]{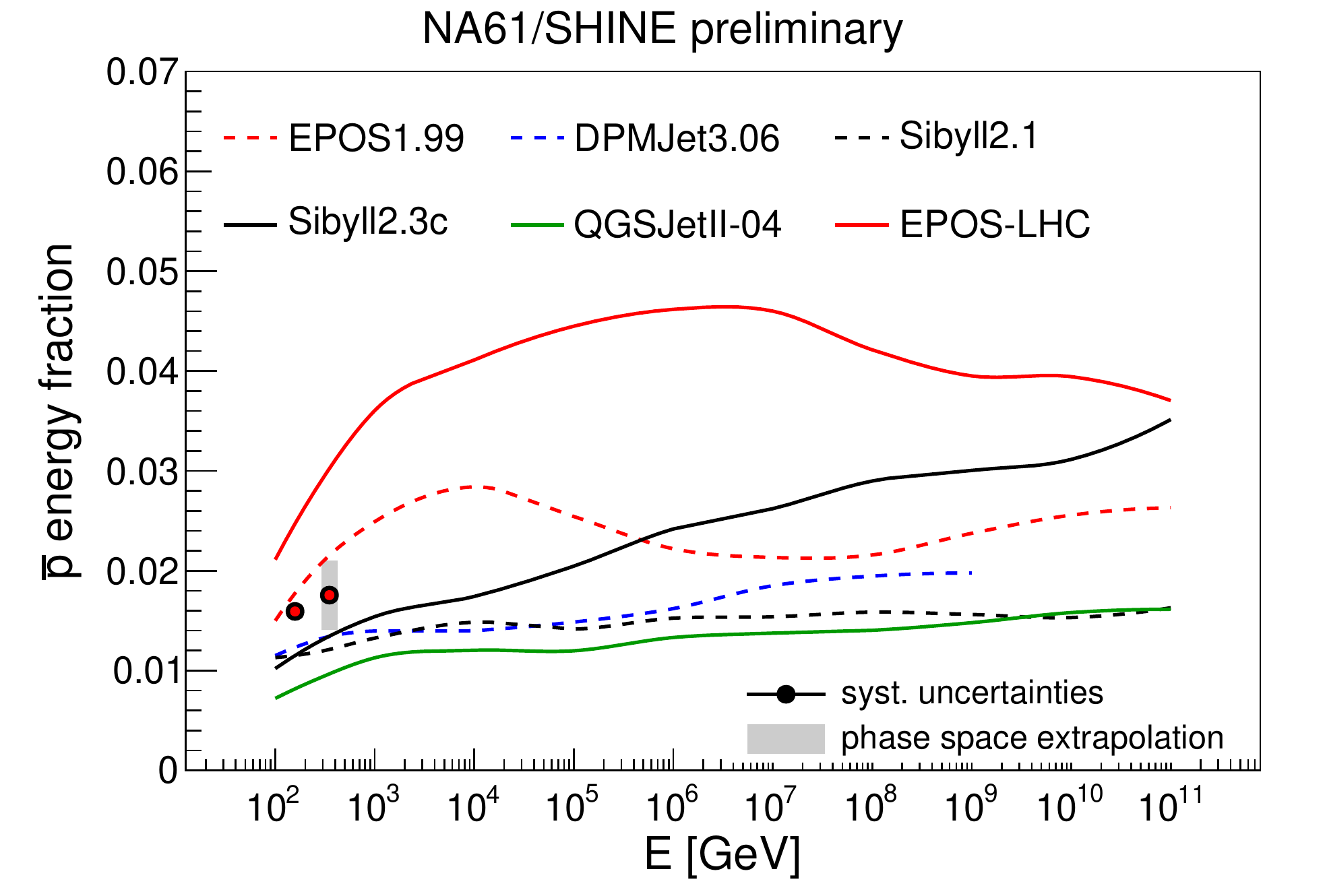}
    \hfill
    \includegraphics[width=0.450\textwidth]{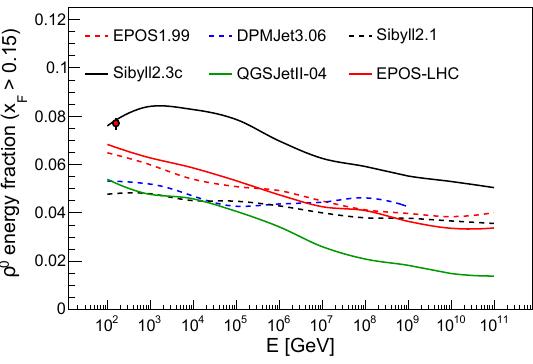}
    \caption{Energy fraction of  anti-protons (left) and $\rho^0$ (right), as measured by NA61/SHINE in $\pi^-$-C interactions, and comparison with predictions from models. (Plot taken from \cite{Unger:2019icr}.)}
    \label{f:NA61}
  \end{center}
\end{figure}

Figure \ref{f:NA61} shows the energy fraction of  anti-protons at 158 GeV/c and 350 GeV/c (left)  and $\rho^0$ at 158 GeV/c (right)  as measured by NA61/SHINE in $\pi^-$-C interactions (\cite{Prado:2017hub,Unger:2019icr} and  \cite{Aduszkiewicz:2017anm} respectively). The comparison with the prediction from models demonstrates the existence of some problems already at low energies.

\begin{figure}[!ht]
  \begin{center}
    \includegraphics[width=\textwidth]{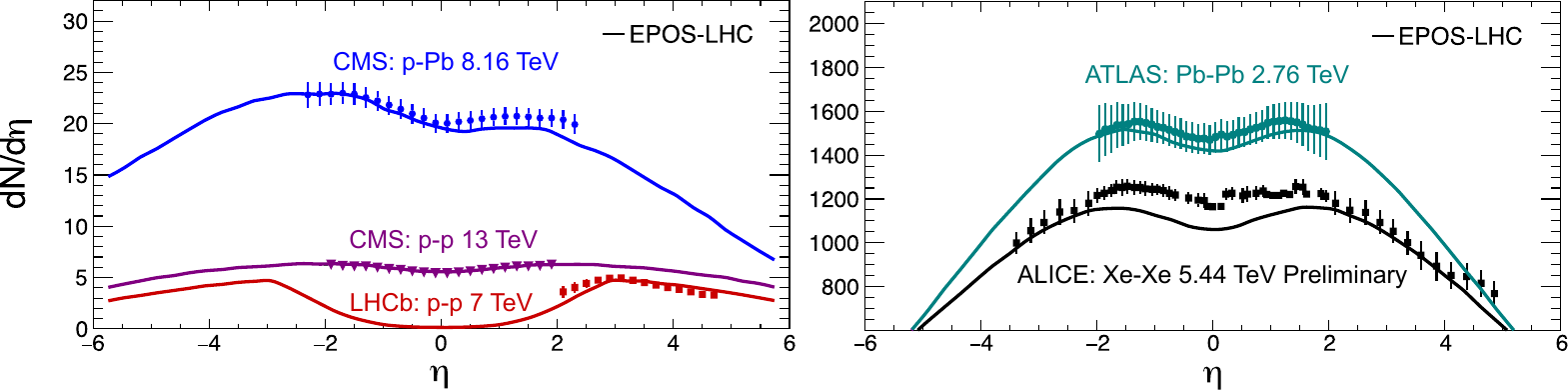}
    \caption{Comparison of $dN/d\eta$ measurements in different colliding systems with EPOS-LHC. (Taken from \cite{Citron:2018lsq}.)
    }
    \label{f:XeXe}
  \end{center}
\end{figure}

Finally, interpolations in models between p-p and p-Pb to other nuclear targets is not straightforward as can be seen in Figure \ref{f:XeXe} for Xe-Xe collisions. A sufficiently accurate theory to predict nuclear modifications for p-Air is not yet available. In \cite{Dembinski:2019icrc} and \cite{Citron:2018lsq} the case was argued for p-O collisions to be studied at LHC with heavy-ion and proton beams.


\section{Experimental Observables}

\begin{figure*}
\centering
\includegraphics[width=0.49\textwidth,clip,trim=30 10 30 10]{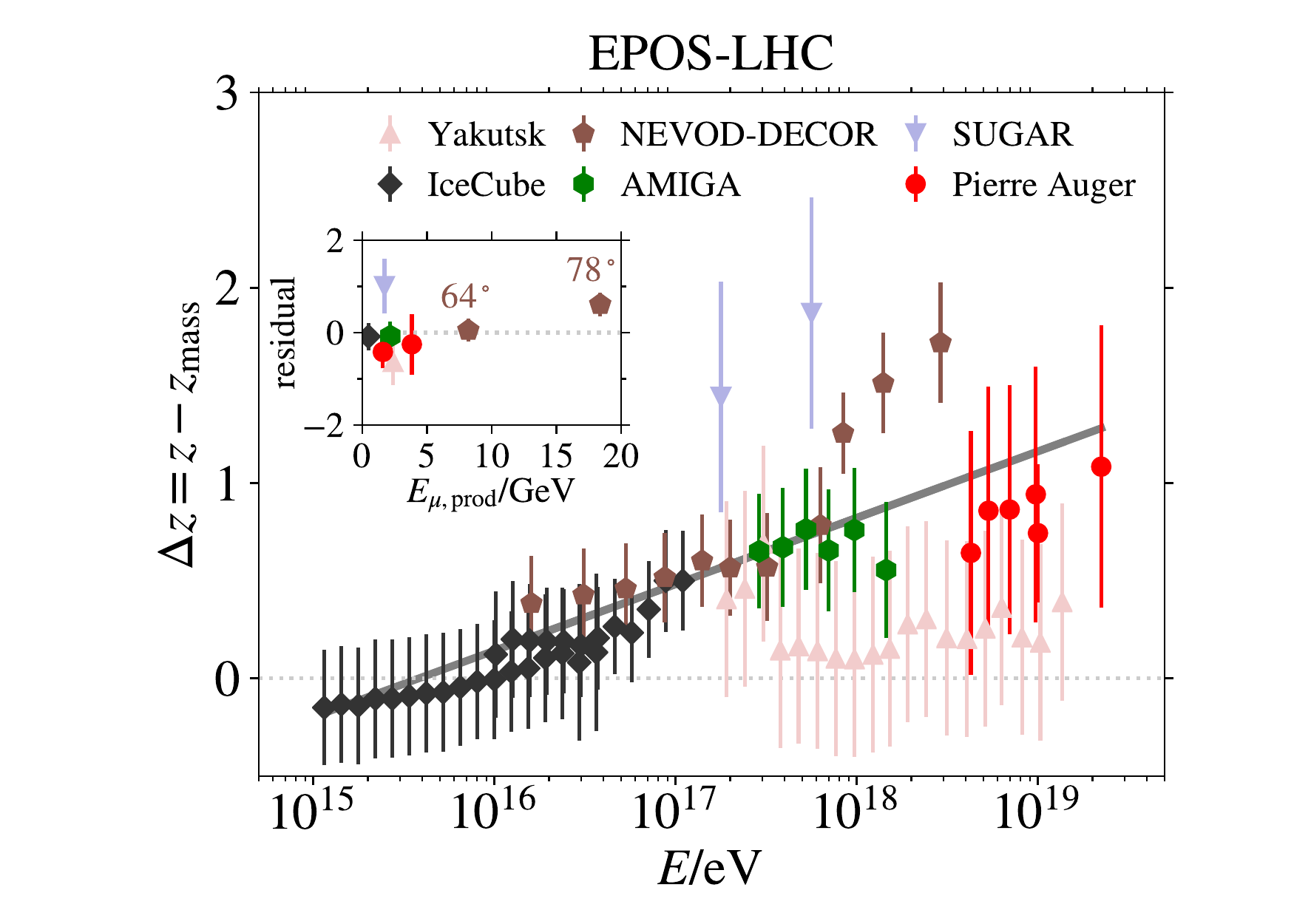}
\hfill
\includegraphics[width=0.49\textwidth,clip,trim=30 10 30 10]{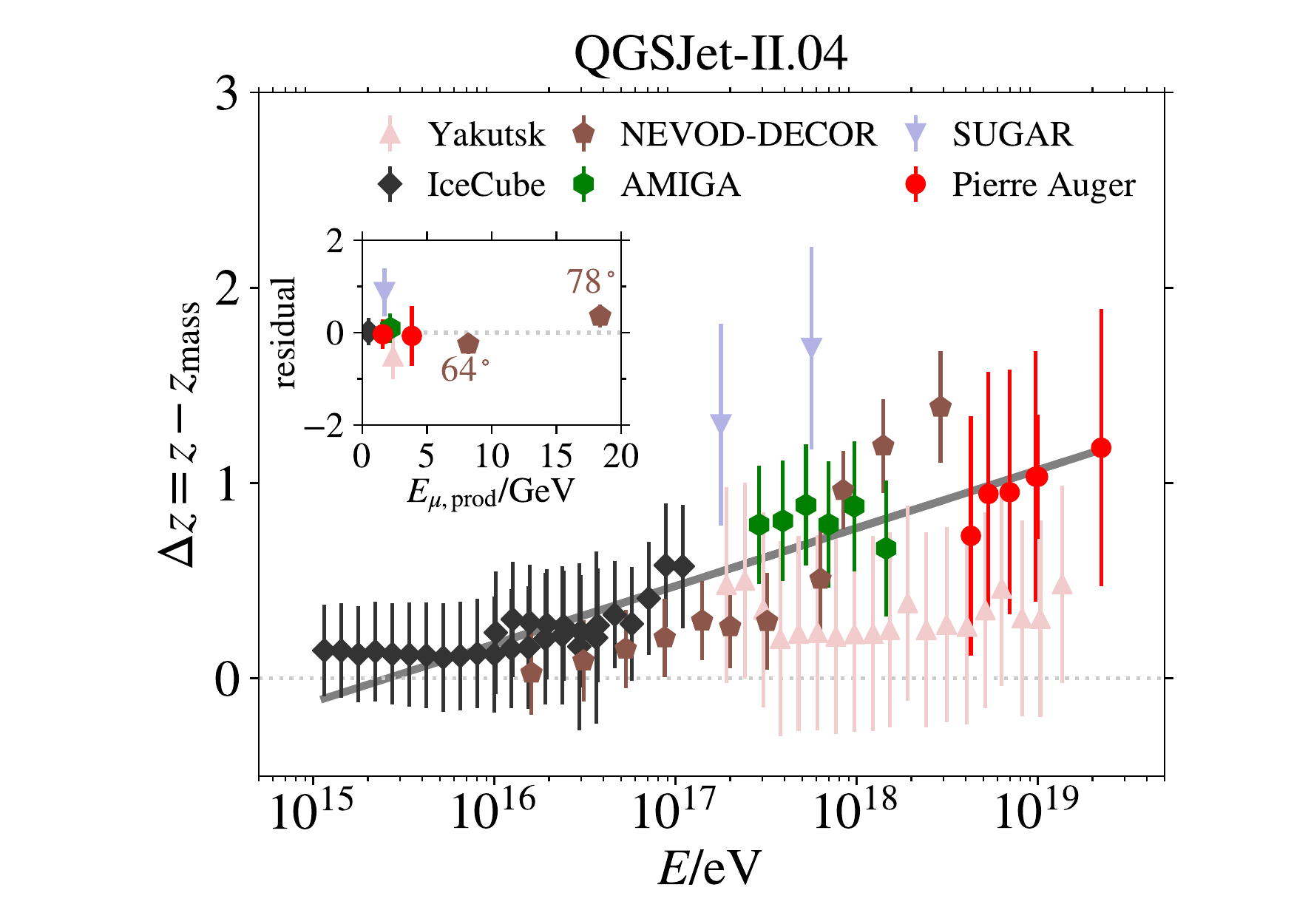}
\caption{$\Delta z=z-z_{\rm mass}$ for EPOS-LHC and QSGJet-II.04.
  The function $\Delta z_{\rm fit} = a + b \log_{10}(E/10^{16}{\rm eV})$ was fitted.
  The inset shows the average residual per data set with respect to the fitted line, $\Delta z-\Delta z_{\rm fit}$,  as a function of the minimum energy of muons at the production point high in the atmosphere, $E_{\mu \, {\rm prod}}$. (Taken from \cite{Cazon:WHISPICRC19}.) 
  }
\label{fig:zfit}
\end{figure*}

After the first report in 2000 by the HiRes/MIA collaboration about a muon deficit in simulations ({\it aka}, muon excess in data) between $10^{17}$ to $10^{18}$\,eV \cite{AbuZayyad:1999xa}, many more experiments have contributed with measurements. NEVOD-DECOR \cite{Bogdanov:2010zz,Bogdanov:2018sfw} observed a muon deficit in simulations starting around $10^{18}$ eV as did the SUGAR array~\cite{Bellido:2018toz}. The Pierre Auger Observatory~\cite{Aab:2014pza,Aab:2016hkv} and Telescope Array~\cite{Abbasi:2018fkz} reported a muon deficit with respect to the latest models in the energy range around and above $10^{19}$ eV.
On the other hand, KASCADE-Grande~\cite{Apel:2017thr} and EAS-MSU~\cite{Fomin:2016kul} reported no discrepancy around $10^{17}$ eV in the muon number.
A comprehensive collection of muon measurements, which also include data from IceCube~\cite{Gonzales:2018IceTop}, AMIGA~\cite{mueller2018}, and unpublished data from Yaktusk~\cite{yakutskpc}, was presented in \cite{Cazon:WHISPICRC19}. 
A  $z$-scale was introduced to plot the ratio of $\nmu$ with respect to proton simulations with a given model, as
$z=\frac{\ln \nmu -\ln \nmu^p}{\ln \nmu^p-\ln \nmu^{Fe}}= k \ln \left(\frac{\nmu}{\nmu^p}\right)$
with $k\simeq 3$. Where $z=0$ corresponds to the number of muons contained in proton showers,  and $z=1$ in iron showers.
 An energy rescaling was applied to all $z$-values in order to obtain a matching energy spectrum between experiments. Afterwards, the difference with respect to expectations from models was calculated as $\Delta=z-z_{\rm mass}$, where $z_{\rm mass}$ was inferred from a compilation of $\Xmax$ data. Results are shown in Figure. \ref{fig:zfit},
 where a growing muon deficit in the simulations can be observed  above  $10^{16}$\,eV. The slope of this increase in $z$ per decade in energy is 0.34 and 0.30 for EPOS-LHC and QGSJet-II.04 respectively, with 8\,$\sigma$ significance.

\subsection{Muon energy and transverse momentum at production}

In \cite{Cazon:WHISPICRC19}, other trends in the deviations between simulations and data were investigated. The minimum energy required at production for a muon to be detected in each experiment ($E_{\mu \, {\rm prod}}$) was calculated, spanning from $\sim 1$ GeV up to $\sim 19$ GeV, reached by NEVOD-DECOR \cite{Bogdanov:2010zz,Bogdanov:2018sfw}.  Unfortunately, current data did not allow any deviation to be claimed for the muon spectrum at production with respect to models, due to the large inhomogeneity and uncertainties of measurements. On the other hand, results by KASCADE-Grande in this conference \cite{KASCADEICRC19}, show that the measured muon number lies between the model predictions for proton and iron nuclei for all models, with a tendency towards heavier primaries as the EAS zenith angle increases.  The results support previous findings of KASCADE-Grande about a problem in the predicted muon attenuation length, and therefore the E-spectrum of muons between 10 PeV and 1 EeV.

Differences in the E-spectrum of muons might arise from possible differences in the ratio of the $\pi^\pm$/K mix in the hadronic cascade, through the effective critical energy of the mix, and by possible differences in the modelling of the  E-spectra of $\pi$ and K in the hadronic interactions. Indeed, \cite{Cazon:2012ti,Cazon:2019nop} and Figure  \ref{f:universality} (right) show departures from universality across models. The E-spectrum of muons  is therefore one of the favourite indirect observables in the shower to constraint hadronic physics. In \cite{Espadanal:2016jse}, the effect of changes in the E-spectrum of muons on different shower observables was studied: namley,  $\Xmumax$ and those related to the distribution of the muons at ground level, that affect the reconstruction of $\nmu$. In general, observables are expected to display a distinctive zenith angle dependence, since the atmosphere imposes different energy thresholds for muons, as described in the previous paragraph.  Variations are expected for  the 2-dimensional distribution of muons at ground level in very inclined showers due to geomagnetic field effects \cite{Billoir:2015cua}. Close to the shower core, where geometrical paths of muons are reduced to a minimum, different arrival times of muons are directly linked to their effective velocities and therefore a direct mapping between time and energy might be used \cite{Cazon:2012ti}.

Finally, there are no publications\footnote{to the best of my knowledge} that demonstrate any deviation of the bulk of the $p_t$-distribution of muons from the universal expectation. Nevertheless, the high $p_t$-tail ($p_t>2$ GeV/c) has been claimed to be poorly described by simulations in \cite{Abbasi:2012kza}.

\subsection{Muon Production Depth}
The Muon Production Depth (MPD) distribution in EAS tracks the longitudinal development of the hadronic cascade, in particular the depth where mesons decayed into muons. It is thus expected to reach its maximum $\Xmumax$ around the depth where the average energy of mesons reaches the effective critical energy of the $\pi^\pm$/K mix. The longer the number of generations in the cascade, the deeper $\Xmumax$ is. It is therefore sensitive to those phenomena which are able to {\it delay} or {\it accelerate} the flow from the hadronic into the EM channel, without necessarily changing the multiplicative process.

\begin{figure*}[!t]
 \centering
 \includegraphics[width=0.43\textwidth,clip,trim=00 10 00 10]{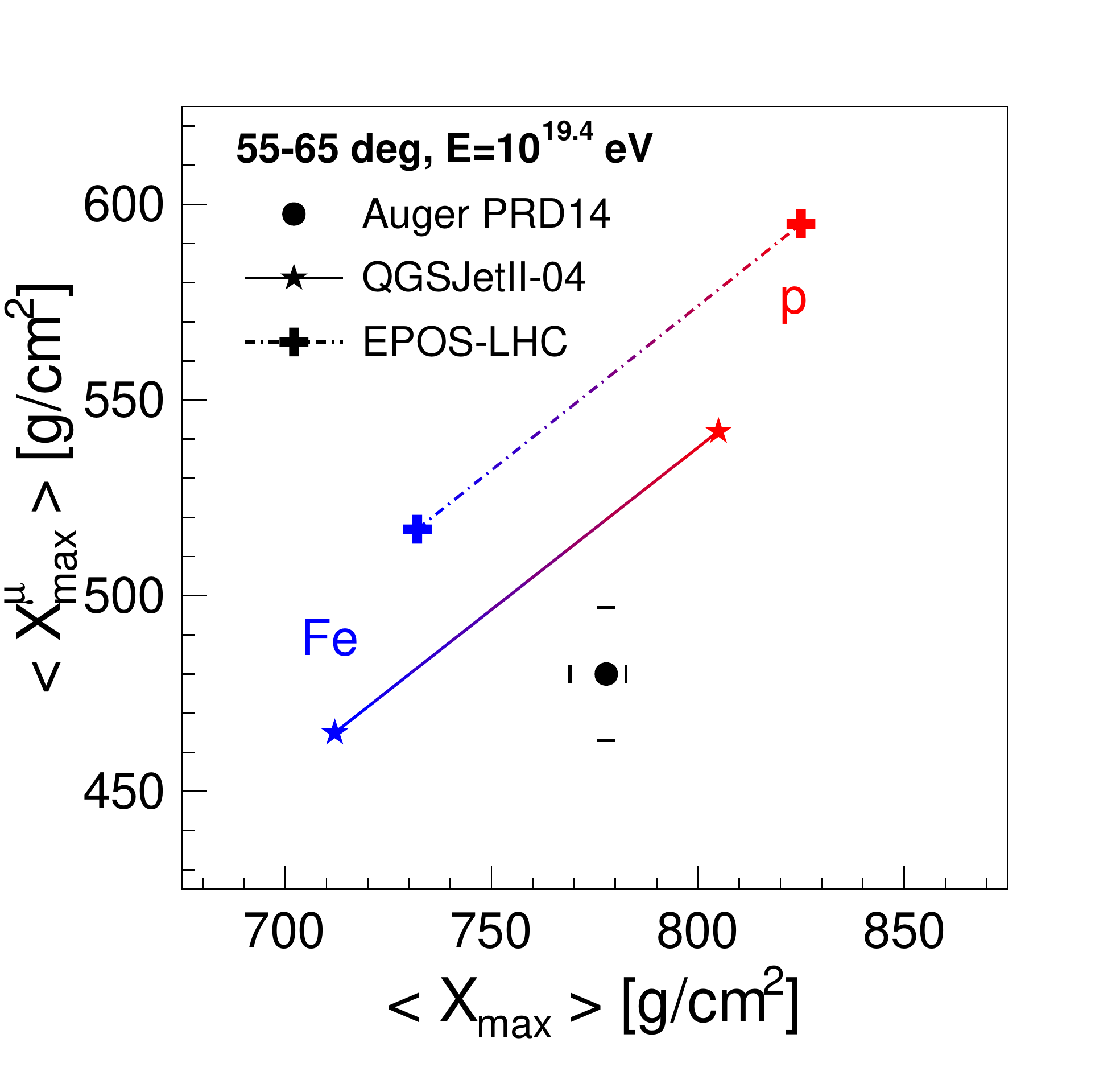}
 \includegraphics[width=0.47\textwidth,clip,trim=00 10 00 10]{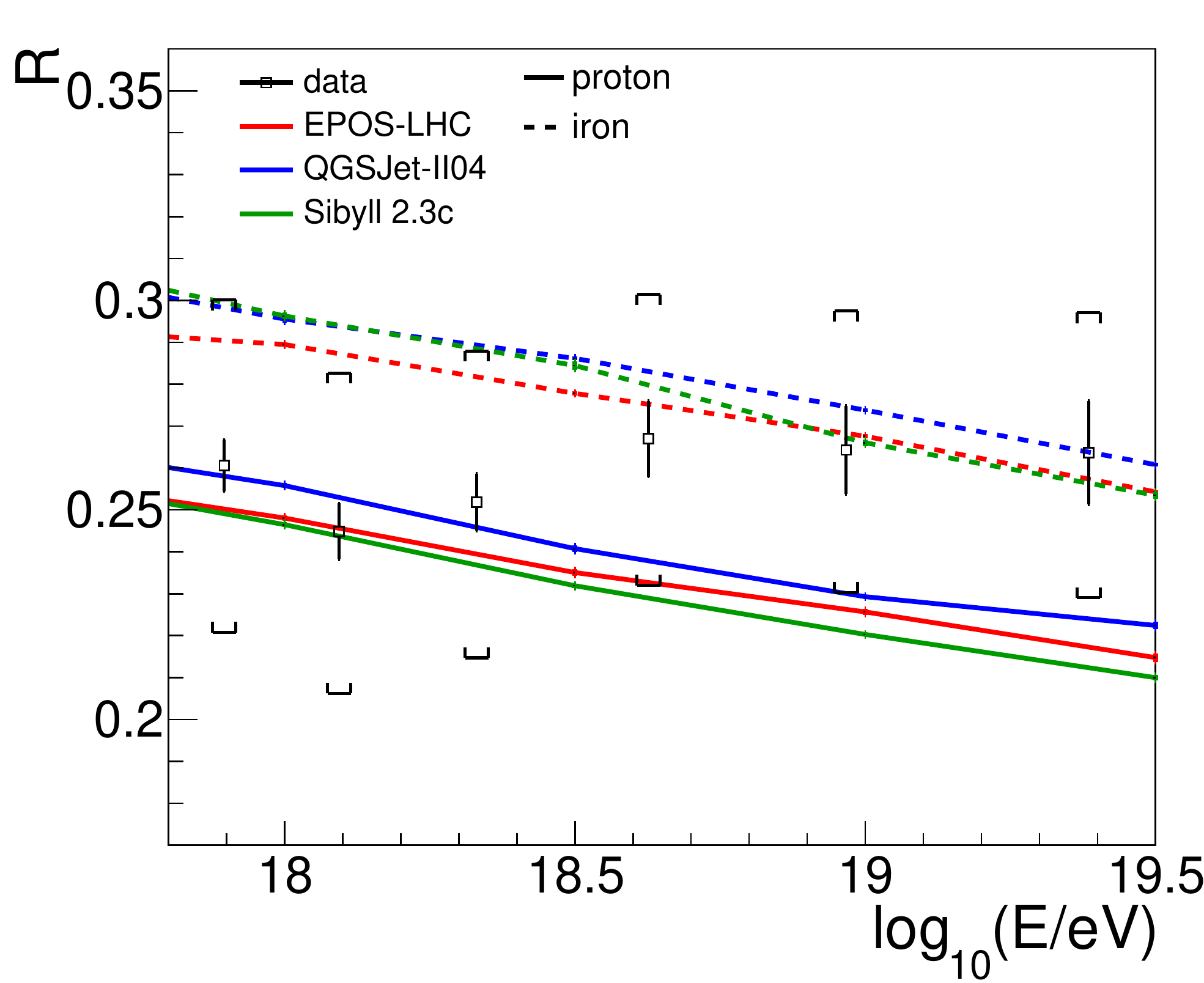}
 \caption{Left: Measurement of $\langle\Xmumax\rangle$  and
   $\langle\Xmax\rangle$ for $10^{19.4}$ eV, with systematics uncertainties,  as well as the phase-space (lines) occupied by EPOS-LHC and QGSJETII.04 models, extracted from \cite{Aab:2014dua}.
     Right: Parameter $R$ of the average electromagnetic profile as a function of the EAS energy as measured by Auger, as well as predictions for different hadronic models, extracted from \cite{Aab:2018jpg}.
 }
 \label{fig:MPDRL}
\end{figure*}

In \cite{Apel:2011zz} the longitudinal development of muon production was investigated at energies between $10^{15}$ and $10^{17}$ eV, showing some tension with the available pre-LHC hadronic models. Later, in \cite{Aab:2014dua} the depth at which the muon production reached a maximum ($\langle \Xmumax \rangle$) was analysed above $10^{19.2}$ eV, for muons which arrived at more than 1700 m from the shower core and for showers between $55^\circ$ and $65^\circ$ zenith angle.
Figure \ref{fig:MPDRL} (left) displays the $\langle\Xmumax\rangle$  and   $\langle\Xmax\rangle$ at $10^{19.4}$, showing  incompatible values for QGSJEtII.04, and
an extremely deep value of $\Xmumax$ for EPOS-LHC, making it lie even out of the p-Fe reach.


In \cite{Ostapchenko:2019ubd} it was demonstrated that a substantial part of the present uncertainties in model predictions for $\Xmumax$ is related to very
high energy pion-air collisions, in particular the inelastic cross sections and the production spectra of mesons and nucleons.
The deep $\Xmumax$ of EPOS-LHC was therefore shown to be caused by a too abundant baryon production in pion air interactions at high energies, and a rather high rate of inelastic diffraction for pion-air collisions \cite{Pierog:2017nes}.
It was shown that such large  differences $\Delta \Xmumax\simeq $50 g cm$^{-2}$ translate into milder differences into the EM profile, $\Delta \Xmax\simeq $ 15 g cm$^{-2}$.
The measurements of MPD provide a unique opportunity to constrain the treatment of pion-air interactions at very high energies and to reduce the uncertainties in $\Xmax$ stemming from models.

\subsection{Electromagnetic Component}

The traditional Gaisser-Hillas can be rewritten as
\begin{equation}
\label{usprl}
\frac{dE}{dX} = \left(1+R \, \frac{X'}{L}\right)^{R^{-2}}
                \exp\left(-\frac{X'}{R \, L}\right),
\end{equation}
where $R=\sqrt{\lambda/|X'_0|}$, $L=\sqrt{|X'_0|\lambda}$ and $X'_0 = X_0 - \Xmax$. In this notation, Equation \ref{usprl} is a Gaussian with standard deviation $L$, multiplied by a term that introduces an asymmetry governed by $R$. 
The parameter $R$ is sensitive to the start-up of the EM shower and is expected to be sensitive to the rate of $\pi^0$ production, whose decay goes into the EM shower. Figure \ref{fig:MPDRL} (right) shows the measurements of $R$ together with model predictions \cite{Aab:2018jpg}. Due to the large systematics uncertainties in this measurement, nothing else can be said other than results are fully compatible with expectations from models.



\section{Discussion}

There have been some attempts to explain the muon problem by increasing the hadronic energy fraction of interactions $f$, like the formation of a Strange Fireball \cite{Anchordoqui:2016oxy}, String Percolation \cite{AlvarezMuniz:2012dd}, Chiral Symmetry Restoration \cite{Farrar:2013sfa},  by increasing the inelastic cross section \cite{Farrar:2019cid}, or for instance resorting to Lorentz Invariance Violation \cite{Aloisio:2014dua} by effectively enlarging the lifetime of $\pi^0$ and keeping them contributing to the hadronic cascade.



The deficit of $\nmu$ with respect to expectations could be produced by small deviations $\delta f$ accumulated along a number $n$ of generations as $\nmu \propto (f+\delta f)^n$.
For instance, a 5\% deviation per generation converts into $\sim 30\%$
deviation after 6 generations.

On the other hand, a large single deviation from expectations ($\delta f
\simeq 0.3-0.6$) would be most likely to occur at the first generation, which
is the one farthest away from the reach of accelerator experiments, with less
direct experimental constraints. A change of the multi-particle production of the 1st interaction with energy would result in a change to
$f_1$. The relation $\frac{\mathrm{d} \ln \nmu}{\mathrm{d} \ln E} = \beta $  then becomes
%
  $\frac{\mathrm{d} \ln \nmu}{\mathrm{d} \ln E} = \frac{\mathrm{d} \ln f_1}{\mathrm{d} \ln E}+\beta $.
%
Hence, a sudden change of $f_1$ would produce a change in logarithmic
slope. The continuous and smooth deviation of simulations with respect to data
from low energies seen in Figure \ref{fig:zfit} supports the hypothesis of a small cumulative,
generation after generation, effect rather than a sudden change.

\begin{figure}
  \includegraphics[width=0.44\textwidth,trim=10 00 25 0]{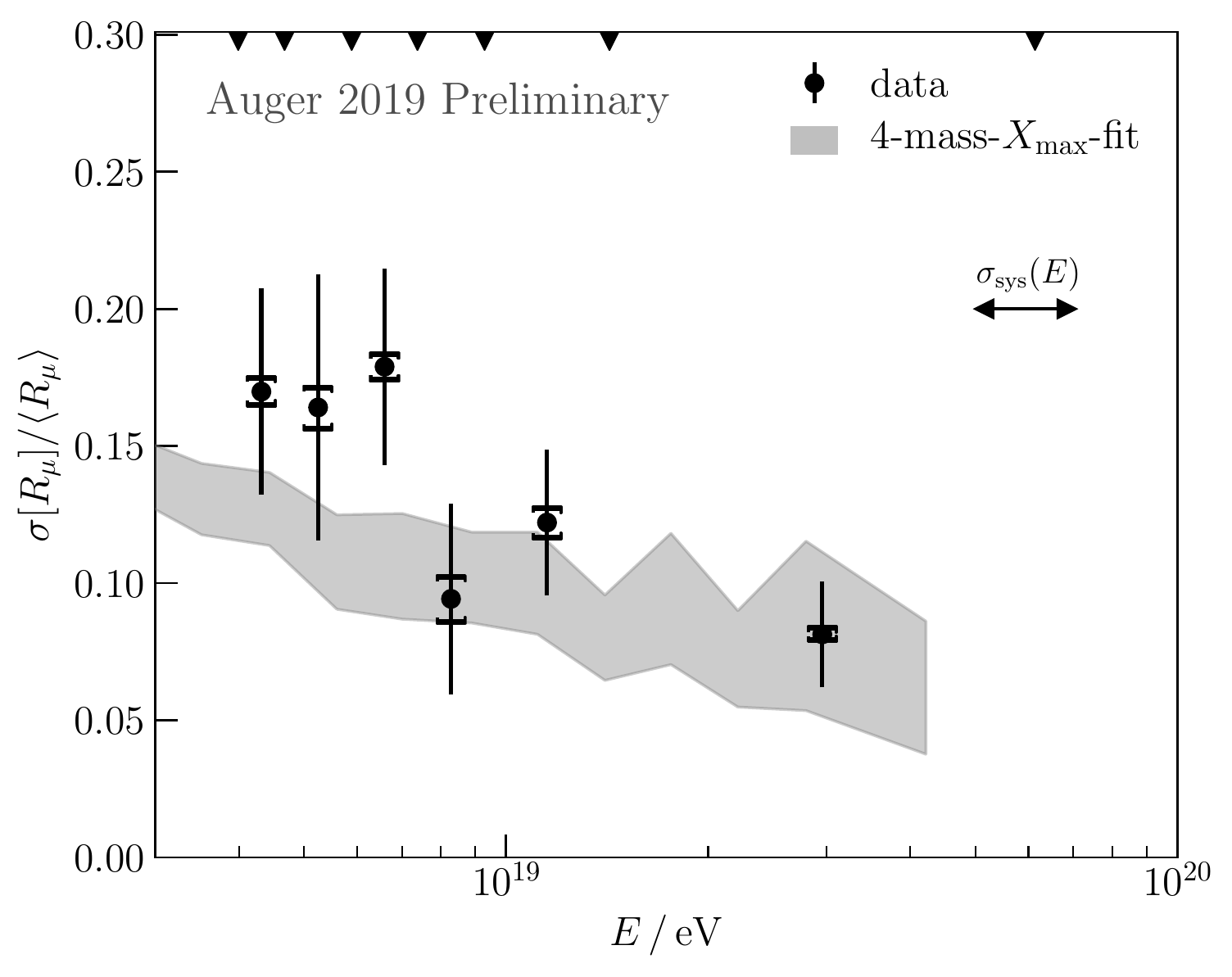}
  \includegraphics[width=0.46\textwidth,trim=10 10 25 0]{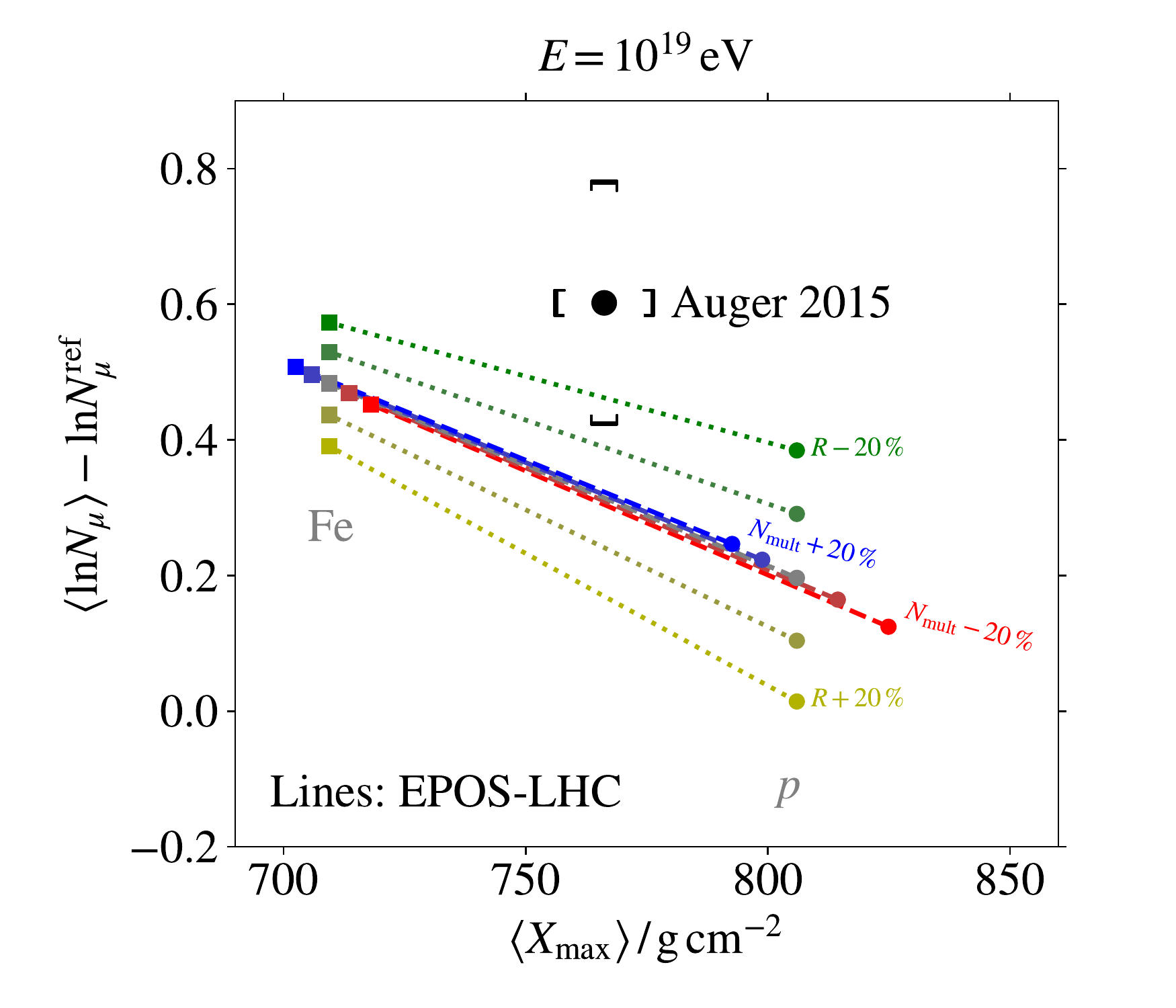}

  \caption{Left: Measurement of the fluctuation of the number of muons by Auger, compared to expectations from composition measurements ($\Xmax$), taken from \cite{Riehn:2019icrc}. Right: Impact of changes of the hadron multiplicity $N_{\rm mult}\equiv m$ (dashed lines) and the energy ratio $R\equiv\frac{f_{\rm EM}}{f}$ (dotted lines) in collisions at the LHC energy of $\sqrt{s}=$13\,TeV on EPOS-LHC predictions for the air shower observables $\xmax$ and $\mlnnmu$ in $10^{19}$\,eV air showers. Data point is from Pierre Auger Observatory~\cite{Aab:2014pza}. The model lines represent all CR-primary mixture from pure proton (bottom right) to pure iron (top left). The dashed and dotted lines represent modifications of $N_{\rm mult}$ and $R$ in steps of 10\% from their nominal values. (Taken from \cite{Baur:2019cpv}.)}
\label{fig:EMfraction}
\end{figure}

%

In \cite{Cazon:2018gww}, a variable of the 1st interaction that combines the multiplicity and energy taken by the particles feeding the hadronic cascade  was defined as
\begin{equation}
  \alpha_1=\sum^m (E_j/E_0)^\beta
\end{equation}
where the index $j$ runs over all particles contributing to the hadronic cascade and  $\beta\simeq 0.9$. The correlation of $\alpha_1$ of the first interaction with the number of muons of the shower is 75\%, whereas the correlation with $f_1$ is slightly smaller, 65\%, due to the diffractive events. For most practical applications one can simply  take  $\alpha_1 \simeq f_1$. The relative fluctuations of $\nmu$ can be expressed by a 
quadratic sum of the relative fluctuations of $\alpha_n$ in the different 
generations $n$ 
%
\begin{equation}
\left( \frac{ \sigma(\nmu)}{\nmu} \right)^2 \simeq
\left(\frac{\sigma(\alpha_1)}{\alpha_1}\right)^2+
\left(\frac{\sigma(\alpha_2)}{\alpha_2}\right)^2+...
+\left(\frac{\sigma(\alpha_{\rm    c})}{\alpha_{\rm c}}\right)^2
  \label{eq:HeitlerFluctuations}
\end{equation}
where $\sigma(\alpha_n) \propto {1}/{\sqrt{m_1 \cdot  m_2 \cdot ... \cdot m_{n-1}}}$, 
which clearly decreases as the generation number gets higher. 
As a result the relative fluctuations of $\nmu$ are dominated by the
fluctuations of $\alpha_1$ in the 1st interaction~\cite{Cazon:2018gww}.  In
p-Air interactions $\sim 70\%$ of the variance is due to the first
interaction, whereas for nuclei of mass $A$, the fluctuations are reduced by a factor $\sim 1/\sqrt{A}$.


In this conference, Auger presented for the first time \cite{Riehn:2019icrc}, the fluctuations of the number of muons as a function of the shower energy, which is shown in Figure \ref{fig:EMfraction} (left), along with expectations from composition analysis. While interaction models give a description of the relative fluctuations which is compatible with data, they show a significant discrepancy with the average muon scale. The values $\langle \nmu \rangle$ and  $\sigma(\nmu)/\langle \nmu \rangle$ depend on different aspects of the shower development.
For $\langle \nmu \rangle$, all generations have an equal contribution~\cite{Matthews:2005sd}, while $\sigma(\nmu)/\langle \nmu \rangle$ is dominated by the first interactions~\cite{Cazon:2018gww} through $\alpha_1$-fluctuations. 



Figure \ref{fig:EMfraction} (right), taken from \cite{Baur:2019cpv}, shows the impact of changing the ratio of electromagnetic to hadronic particles,  $R \equiv \frac{f_{\rm EM}}{f}=(f^{-1}-1)$ in the observables $\nmu$ and $\xmax$ together with a measurement by Auger \cite{Aab:2014pza}, by following the methodology described in \cite{Aab:2014pza}. When $m$ ($N_{\rm mult}$) is modified the simulated line shifts parallel to itself: the multiplicity has a correlated effect on $\xmax$ and $\lnnmu$. On the other hand modifying $R$ changes the muon number and leaves $\xmax$ unchanged. A decrease of $R$ of 15\% at the $\sqrt{13}$ TeV would be enough to make simulations compatible with air shower data at $10^{19}$ eV. 
In \cite{Baur:2019cpv}, $R$ was proposed as an experimental observable  to be measured in LHC calorimeters as a function of pseudorapidity and central charged particle multiplicity.
It is claimed to be a new handle for the explanation of collective hadronisation in p-p collisions, and distinguish between quark-gluon-plasma-like (QGP-like) effects, from alternative more microscopic effects. Precision measurements of $R$ to 5\% at the LHC could contribute to a better understanding of muon production in air showers \cite{Citron:2018lsq}.

\section {Direct Measurement}

\begin{figure}[h]
  \centering
  \includegraphics[width=0.45\textwidth,clip]{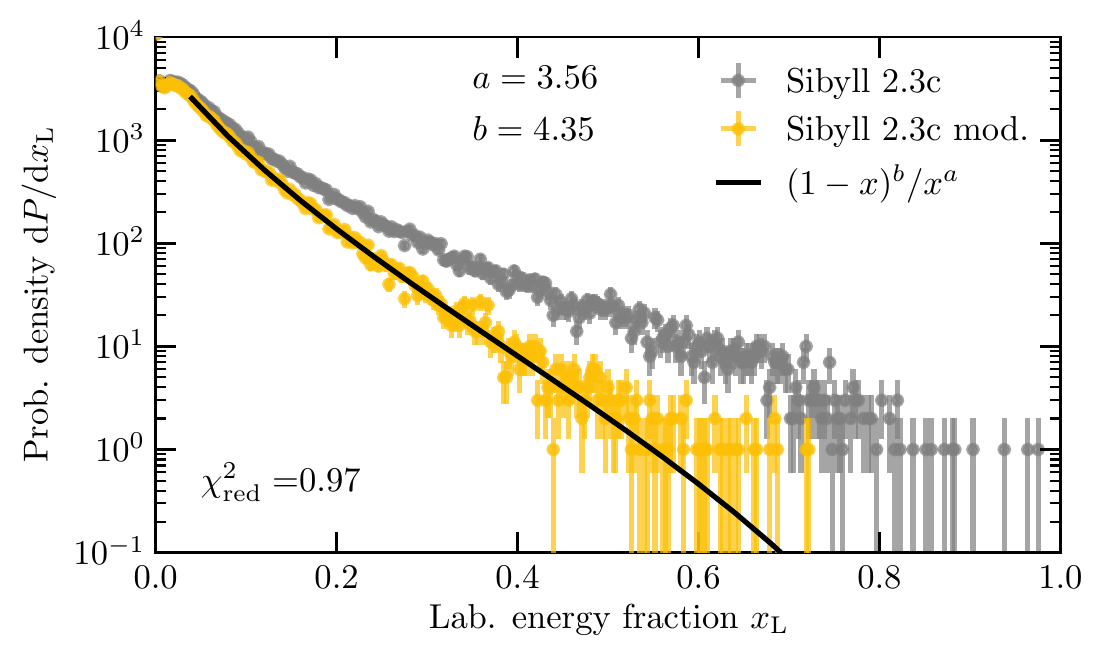}
   \includegraphics[width=0.45\textwidth,clip]{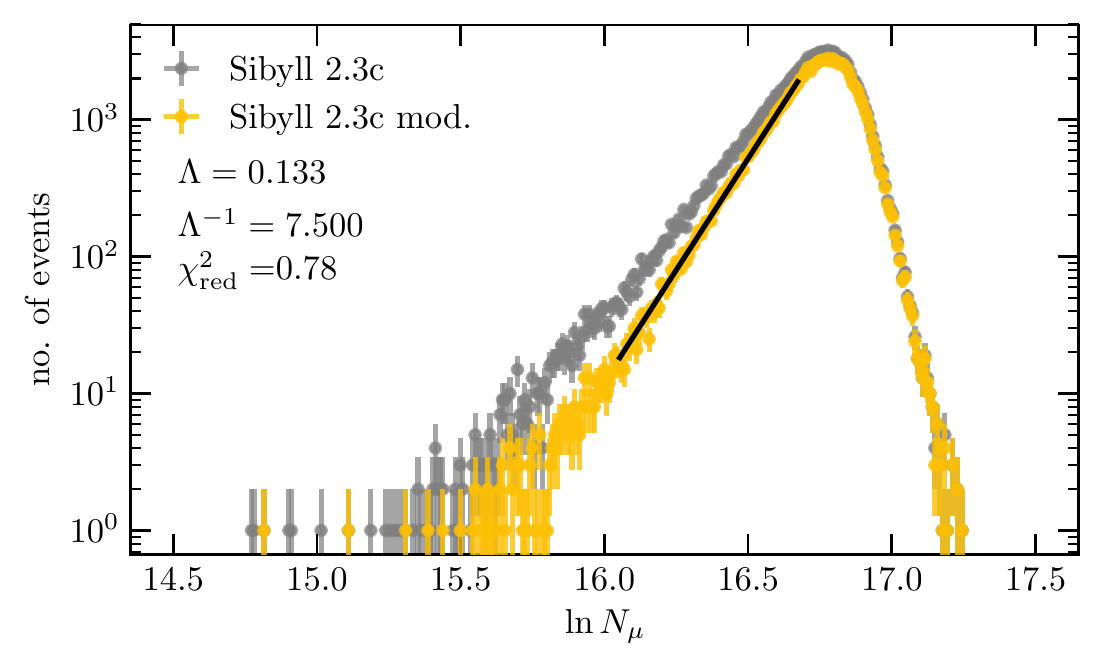}
  \caption{Left: Inclusive production cross-section of $\pi^0$ as a function of lab.\ energy for SIBYLL-2.3c~\cite{Riehn:2017mfm}. A suppressed cross-section at large $x_{\rm L}$ is shown in yellow. Right: Distribution of the number of muons at ground level for the two production spectra shown in the figure on the left. (Taken from \cite{Conceicao:2019icrc}.)}
  \label{fig-reshape}
\end{figure}

The measurement of the p-Air cross sections by cosmic ray measurements was the only case where one could directly perform a measurement on the 1st interaction. It relied on the fact that p-Air interactions will produce a detectable deep tail on the  $\Xmax$ distributions, which is a direct mapping of the depth of the first interaction $X_0$ \cite{Collaboration:2012wt,Ulrich:2015yoo}. In this conference, a method was presented to measure the high energy end of the $\pi^0$ spectrum \cite{Conceicao:2019icrc}. It relies on the fact that extreme low-$\nmu$ fluctuations on p-Air interactions are visible provided there are enough protons in the UHECR composition. Figure \ref{fig-reshape} displays the effect of changing the energy spectrum of $\pi^0$ in the first interaction (only), and its measurable effects on the extreme low-$\nmu$ fluctuations of the p-Air interactions.

\section{Conclusions}

Ultra High Energy Cosmic Rays present a great opportunity to explore particle physics beyond the reach of accelerators. Our understanding of hadronic physics in the forward region and at the highest energies is a mere extrapolation to the unknown, and therefore is subject to uncertainties. The mass interpretation of UHECR, on the other hand, needs EAS simulations which make use of the high energy hadronic models. The UHECR mass inference inherits the model uncertainties, hampering the quest for their UHECR origin.

 LHC measurements have improved the agreement of models in describing  p-p collisions, but important differences remain in describing the p-Air and specially the $\pi$-Air collisions, which dominate the development of the EAS hadronic cascade. New measurements for p-O are being proposed for the new LHC phase, to fill the gap of intermediate nuclei measurement at the highest energies.

 Muon EAS observables, like the Muon Production Depth (MPD), the muon number, or the muon energy spectrum are showing some interesting facts:
 the MPD results are demonstrated to be quite sensitive to the modelling of the difractive $\pi$-Air interactions, and can indirectly reduce the $\Xmax$ model uncertainties.
 The E-spectrum of muons is sensitive to the energy spectrum of mesons, and to the ratio of $\pi$/K mix of the hadronic cascade.
 Finally, the muon deficit in simulations ({\it aka} muon excess in data) has been shown to start around $10^{16}$ eV with a smooth steady trend. The shower-to-shower fluctuations of the muon number are dominated by the fluctuations of the partition of energy in the first interaction. The first measured results of the muon number fluctuations around $10^{19}$ eV disfavour a large departure from expectations in the first UHECR-Air interactions. Therefore, the muon deficit should come from an accumulation of small deviations along the different generations of the hadronic cascade, in  $meson$-Air or nucleon-Air interactions. There are several proposals to increase the hadronic energy fraction, (\cite{Pierog:2019icrc,Anchordoqui:2016oxy,Farrar:2019cid,Farrar:2013sfa}), which need to be tested against the rest of EAS observables.

 Finally, a measurement of the low tail of the muon number fluctuations was proposed, as it maps the inclusive production $\pi^0$ cross-section in the first p-Air interaction, in a similar way that the high $\Xmax$-tail was used to measure the p-Air cross section. This would be the first measurement of a multi-particle production characteristic beyond the 100 TeV scale.

\section*{Acknowledgments}
LC wants to thank F. Riehn for having helped in the production of new figures in this manuscript. Also, to  R. Concei\c c\~ ao, M. A. Martins, H. Dembinski and F. Riehn for numerous discussions and R. Clay and S. Andringa for a careful reading of this manuscript.  
LC acknowledges funding by Funda\c c\~ ao para
a Ci\^ encia e Tecnolog\' \i a, COMPETE, QREN, and European
Social Fund.

\end{document}